\documentclass[%
 reprint,
 amsmath,amssymb,
 aps,
]{revtex4-2}
\usepackage{graphicx}% Include figure files
\usepackage{dcolumn}% Align table columns on decimal point
\usepackage{bm}% bold math
\usepackage{xcolor}
\usepackage{hyperref}% add hypertext capabilities
\usepackage{braket}%FOR KET and BRA notation
\begin{document}

\preprint{APS/123-QED}

\title{Chiral-induced spin selectivity augments quantum coherence in avian compass%Coherence in a Chiral induced spin selective (CISS) environment
}% Force line breaks with \\

\author{Yash Tiwari}
% \altaffiliation[Also at ]{Department of Electronics and Communication,Indian Institute of Technology, Roorkee, Uttrakhand 247667, India}%Lines break automatically or can be forced with 

\author{Vishvendra Singh Poonia}%
 \email{vishvendra@ece.iitr.ac.in}
\affiliation{%
Department of Electronics and Communication,Indian
Institute of Technology, Roorkee, India %\textbackslash\textbackslash
}%


%\affiliation{
%\affiliation{
%\affiliation{%


\date{\today}% It is always \today, today,

\begin{abstract}
 
This work investigates the effect of chiral-induced spin selectivity (CISS) on quantum spin coherence in the radical pair mechanism of avian magnetoreception. %We examine the impact of CISS on the avian compass coherence. 
Additionally, we examine the utilitarian role of coherence for the avian compass by analyzing its correlation with the yield of the signaling state. %We take the help of the following two coherence quantifiers 1) relative entropy of global coherence, 2)relative entropy of local coherence. 
%We have done this analysis primarily for six-nuclei cryptochrome based on the radical pair model. 
We find that both the relative entropy of global coherence and local coherence in the radical pair increases with CISS. However only the global coherence exhbit the utilitarian role for the avian compass.%   an increase in coherence for both the quantifiers used in our study. 
We also analyze the interplay of dipolar interaction with the CISS and their effect on coherence of the radical pair. %Further, we also consider the dipolar and exchange interactions in this study and observe their result along with CISS for six nuclei cases. 
Further, we analyze the effect of environmental decoherence along with CISS.%  is followed by the effect of decoherence on system coherence. %We then investigate the utilitarian role of coherence in the CISS-assisted-avian compass. 
We conclude that a high CISS results in a high correlation of global coherence with signaling state yield. %Thus a high CISS results in higher global coherence, and due to high correlation results in a high yield of signaling state. 
It confirms that CISS plays an important role both for compass sensitivity and coherence in the avian compass.

\end{abstract}

\maketitle

%\tableofcontents

\section{\label{INTRODUCTION}Introduction}
Coherence is a resource for quantum technologies, and its existence and utility for the biological processes occurring at ambient conditions have been extremely intriguing.
%has been considered a resource of quantum biology playing a significant role in the process like photosynthesis \cite{blankenship2021molecular}, sense of smell \cite{turin1996spectroscopic,franco2011molecular} and avian magneto-reception \cite{hore2016radical}. 
Avian magnetoreception is one such biological process where the radical pair spin dynamics, along with the role and utility of quantum coherence, has been investigated from several aspects \cite{cai2013chemical,kominis2020quantum,jain2021avian,gauger2011sustained,ritz2000model,hore2016radical,smith2022observations,smith2022driven}.
%a bird's sense of using the earth's magnetic field to find navigational direction.
The radical pair mechanism is based on a spin-sensitive chemical reaction that is mediated by a protein molecule ~\cite{hiscock2018long,hiscock2016quantum,rodgers2009chemical}. Owing to the chirality of protein molecules, the chiral-induced spin selectivity (CISS) effect could play an essential role in the electron transport part of the reaction. The origin of CISS is attributed to the spin-orbit interaction and the electrostatic potential provided by the chiral molecules~\cite{dalum2019theory,michaeli2019origin,matityahu2016spin,gohler2011spin,naaman2012chiral,naaman2015spintronics,xu2021magnetic,wong2021cryptochrome}. It was shown by Fay et al. ~\cite{fay2021chirality} that chirality in conjunction with spin-orbit reaction in electron transfer reaction generates coherence locally. It was done for the electron spin echo experiment.    It was also shown in ~\cite{luo2021chiral} that the prerequisite for forming a radical pair for avian magneto-reception is the transfer of electrons. This transfer or transport of electrons between donor and acceptor occurs in a chiral molecule contributing toward CISS and affecting the operation of the avian compass. 
%This work focus on the avian magneto-reception biological system

Coherence quantifiers are based on the non-diagonal/coherence element of the density matrix of a quantum state~\cite{baumgratz2014quantifying,streltsov2017colloquium,winter2016operational}. It has been suggested in references~\cite{cai2013chemical,katsoprinakis2010coherent} that global coherence rather than local or electronic coherence might enhance the compass sensitivity. Therefore, we correlate the yield of the spin-selective chemical reaction with coherence in a chiral medium for the avian-magnetoreception.

 %\textcolor{red}{A correlation coefficient was used to establish a relation between sensitivity and coherence in \cite{smith2022observations}. However, the study did not address the chirality of the medium.}
 %\textcolor{red}{.}
%Recently it was observed that a chiral molecule acts as a spin filter allowing the transport of only one type of spin~\cite{ray1999asymmetric}. 

In this work, we make use of relative entropy of coherence \cite{baumgratz2014quantifying} and total coherence \cite{jain2021avian} as coherence quantifiers to answer the following questions: i) how does the CISS affect the local and global coherence in avian compass, ii) how do dipolar and exchange interaction in conjunction with CISS affect the total local and global coherence measures,
iii) how does the environmental decoherence affect the multi-nuclei radical pair mechanism, and more importantly iv) does the quantum coherence play any utilitarian role for the avian compass? If yes, in what form? We have considered the case of three nuclei each on flavin adenine dinucleotide FAD and tryptophan TrpH radicals. The FAD act as a donor entity, whereas TrpH act as an acceptor entity. The hyperfine interaction values of these nuclei have been taken from ref.~\cite{hiscock2018long}.  

The manuscript has been organized as follows: Section~\ref{METHODOLOGY} discusses the methodology followed for analysis. Section~\ref{RESULTS}  discusses the results, wherein subsection~\ref{Effect_CISS_Coherence} discusses the effect of CISS on quantum coherence, and subsection~\ref{CISSCOHRENCEEE} explores the impact of electron-electron interactions on system sensitivity along with CISS. Section ~\ref{DIFFRENT_RATE_COHRENCE} illustrate an increase of coherence due to CISS at various rate combinations. Section~\ref{Environmental_Decoherence} demonstrates the effect of environmental decoherence on the system. Section~\ref{Utilitarian} examines the utilitarian aspect of the quantum coherence in the avian compass.

\section{\label{METHODOLOGY}Methodology}
%We start with donor and acceptor molecule located on the retina of the bird. When a sunlight fall on retina, the electrons in the ground state, it get excited to higher energy states leaving a vacancy in the ground state. The donor electron donates the electron to fill this vacancy. Hence two isolated electron is formed one in the ground state of donor(spin operator $S_D$) and other in the excited state of acceptor (spin operator $S_A$), thus forming the radical pair.For more detail please refer to 

In the radical pair model of the avian compass, an electron is photo-excited in the acceptor molecule, creating a vacancy in the ground state. Another electron from a neighboring donor molecule travels in the chiral medium to fill this vacancy. It results in the formation of a radical pair where the spin state of the electron on the donor molecule is $\hat{S}_{D}$ and on the acceptor molecule is $\hat{S}_{A}$~.\cite{luo2021chiral, https://doi.org/10.48550/arxiv.2209.00736}.

The spin state of the above formed radical is governed by Hamiltonian given by~\cite{cintolesi2003anisotropic,fay2020quantum,luo2021chiral}
% =======
% EQ. 1
% =======
\begin{equation} 
\label{Hamiltonian_RPM}
\begin{split}
\hat{H}  = \omega.(\hat{S}_{A}+\hat{S}_{D}) + \sum_{i\in{D,A}}\sum_{k}\hat{S}_{i}.A_{ik}.\hat{I}_{ik} \\ -
J(2\hat{S}_{A}.\hat{S}_{D} + 0.5) +
\hat{S}_{A}.D.\hat{S}_{B}
\end{split}
\end{equation}
where $\omega=g\Bar{\mu_B} \Bar{B}$,  $\Bar{B} = B_0{((cos\theta cos\phi)\Bar{x}+(cos\theta sin\phi)\Bar{y}+(cos\theta)\Bar{z}})$. $B_0$ corresponds to the earth's magnetic field. $\theta$ and $\phi$ correspond to the orientation of the magnetic field with respect to hyperfine tensor~\cite{gauger2011sustained}. $J$ and $D$ are the exchange and dipolar interactions. $A$ is the hyperfine tensor depicting interactions between electrons and neighboring nuclear spins.
% =======
% FIG. 0
% =======
\begin{figure}[htbp]
\centering
\includegraphics[width=90mm,keepaspectratio]{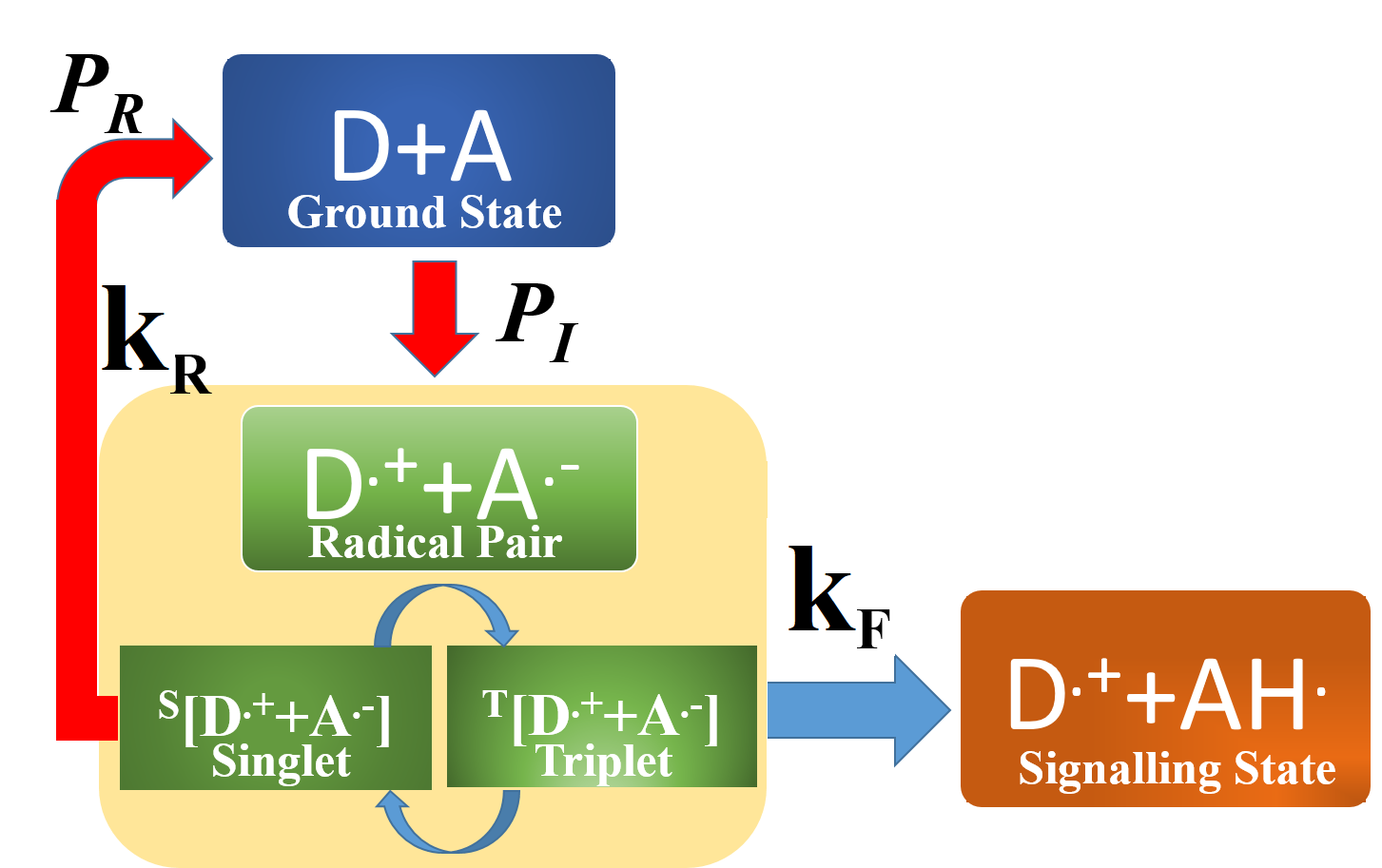}
\caption{The schematic for the CISS-assisted radical pair mechanism where $D$ denotes the donor molecule, $A$ represents the acceptor molecule. $D^{.+}$ is the donor radical, $A^{.-}$ is the acceptor radical, $k_F$ is the protonation rate to the signalling state and $k_R$ is the recombination rate to the ground state. The red arrows represent the role of CISS in the reaction pathways.}
\label{RECOMBINATION}
\end{figure}

The spin state of the radical pair evolves under Zeeman and hyperfine interactions. Along with this evolution of the spin state, the radical pair also recombines back as shown in Fig.~\ref{RECOMBINATION}. The recombination either happens back to the ground state or to the signaling state (via protonation with $H^+$ of the acceptor radical, cf. Fig.~\ref{RECOMBINATION}). The CISS plays role in the formation and recombination of the radical pair as ther involve electron transport through the chiral medium. Therefore, the effect of CISS is captured by the initial state $P_I$ and recombination state $P_R$, also shown with red arrows in Fig.~\ref{RECOMBINATION}. The signaling state does not involve the transfer of electrons (only $H^+$ involved), therefore, CISS is not involved in its formation (shown with the blue arrow in Fig.~\ref{RECOMBINATION}). We define~\cite{luo2021chiral}:

% =======
% EQ. 2
% =======
\begin{equation} 
\label{INITIAL_STATE}
\begin{split}
\ket{\psi_I}=\frac{1}{\sqrt{2}}[\sin(0.5\chi)+\cos(0.5\chi)]\ket{\uparrow_D\downarrow_A}+\\
\frac{1}{\sqrt{2}}[\sin(0.5\chi)-\cos(0.5\chi)]\ket{\downarrow_D\uparrow_A}
\end{split}
\end{equation}
Them the initial density matrix is given as: $P_I= \ket{\psi_I}\bra{\psi_I}\otimes \frac{I}{Z}$, where $\frac{I}{Z}$ corresponds to the mixed state of nuclei, and $Z$ is size of the Hilbert space of the nuclei. The recombination operator $P_R= {\ket{\psi_R}\bra{\psi_R}}$ accounts for recombination to the ground state where $\ket{\psi_R}$ is:
% =======
% EQ. 3
% =======
\begin{equation} 
\label{RECOMBINATION_STATE}
\begin{split}
\ket{\psi_R}=-\frac{1}{\sqrt{2}}[\sin(0.5\chi)-\cos(0.5\chi)]\ket{\uparrow_D\downarrow_A}-\\
\frac{1}{\sqrt{2}}[\sin(0.5\chi)+\cos(0.5\chi)]\ket{\downarrow_D\uparrow_A}
\end{split}
\end{equation}

The CISS parameter $\chi\in [0,\frac{\pi}{2}]$ depends on the spin selectivity of the reactants medium; $\chi = 0$ corresponding to no CISS and $\chi = \pi/2$ correspoding to the maximum CISS. The master equation governing the state evolution of the system is given as: %Eq.\ref{MASTER_EQUATION}.
% EQ. 4
% =======
\begin{equation} 
\label{MASTER_EQUATION}
\begin{split}
\frac{d\hat{\rho}}{dt} =-i[\hat{H},\hat{\rho}(t)]-\frac{1}{2}k_R[P_R,\hat{\rho}(t)]-k_F\hat{\rho}(t)
\end{split}
\end{equation}
 Where $k_F$ is the protonation rate to the signalling state, and $k_R$ is the recombination rate (back to the ground state)~\cite{luo2021chiral,https://doi.org/10.48550/arxiv.2209.00736}.

\section{\label{RESULTS}Results}
This section is divided into three subsections. In the first subsection, we analyze the effect of CISS on the local and global coherence in the radical pair system. Interestingly, we observe that the CISS enhances both local and global coherences in the radical pair system. In the second subsection, we examine the effect of the dipolar  interactions on global and local coherence along with CISS in the RP model. In the last subsection, we study the coherence in RP system as a function of recombination and protonation rates ($k_R$ and $k_F$). 

\subsection{\label{Effect_CISS_Coherence}Effect of CISS on Coherence}
To quantify the coherence in the radical pair system, we use the von-Neumann entropy $S(\rho)$, which is given in Eq.~\ref{ENTROPY} where $Tr$ corresponds to the trace of a matrix. It has a minimum value of zero for pure states and a maximum value of $ln(d)$, where $d$ is the dimension of the Hilbert space of the system. The maximum value corresponds to the maximally mixed state of the system.
% =======
% EQ. 1
% =======
\begin{equation} 
\label{ENTROPY}
\begin{split}
S(\rho)= -Tr(\rho ln(\rho))
\end{split}
\end{equation}
%====================
With the von-Neumann entropy, the coherence quantifier of the RP system can be defined by the relative entropy of local and global coherence, as given in Eq.~\ref{COHRENCE_Local} and Eq.~\ref{COHRENCE_Global}  respectively~\cite{baumgratz2014quantifying}.
% =======
% EQ. 2
% =======
\begin{equation} 
\label{COHRENCE_Local}
\begin{split}
C_{L}(\rho)= S(\rho^{el}_{diag})-S(\rho^{el})
\end{split}
\end{equation}
%====================
% =======
% EQ. 2
% =======
\begin{equation} 
\label{COHRENCE_Global}
\begin{split}
C_{G}(\rho)= S(\rho_{diag})-S(\rho)
\end{split}
\end{equation}
%====================

The local coherence only accounts for the coherence in the electron pair system while the global coherence is the measure of the electron-nuclear coherence. Therefore, in Eq.~\ref{COHRENCE_Local}, $\rho^{el}$ is the density matrix of the electrons that is obtained after partial trace of $\rho(t)$ (obtained from Eq.~\ref{MASTER_EQUATION}) over the nuclear spin subspace. $\rho_{diag}^{el}$ is the density matrix of the electron pair without the off-diagonal terms. In Eq~\ref{COHRENCE_Global}, $\rho$ is the density matrix of the combined (electrons + nuclei) system, and $\rho_{diag}$ is the combined system's density matrix without the off-diagonal terms. We also use a quantifier called the total coherence measure defined in~\cite{jain2021avian} that captures the coherence summed over the entire evolution period. It is given as:
% =======
% EQ. 2
% =======
\begin{equation} 
\label{TOTALCOHRENCE}
\begin{split}
M_{i}(\rho)= \int_{0}^{\infty} C_{i}(\rho(t)) \,dt
\end{split}
\end{equation}
%====================
Here $i \in\lbrace L, G\rbrace$, corresponds to the local and global coherence respectively.
% =======
% FIG. 0
% =======
\begin{figure}[htbp]
\centering
\includegraphics[width=90mm,keepaspectratio]{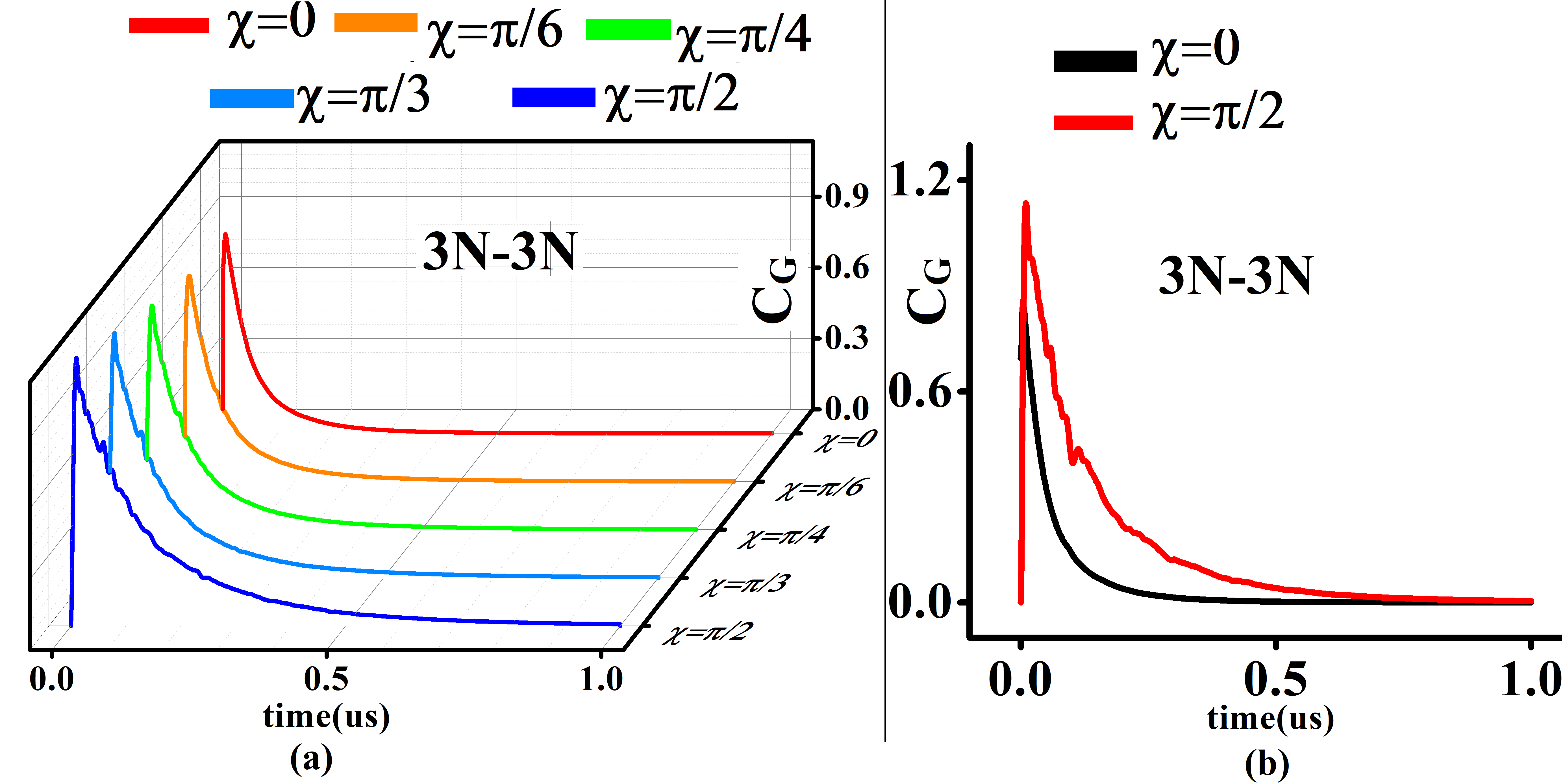}
\caption{(a) Relative entropy of global coherence $C_G$ at $(k_F, k_R) = (10^6, 10^8)~s^{-1}$ for five distinct values of $\chi$ corresponding to varying degree of spin selectivity due to CISS ($0$, $\frac{\pi}{6}$, $\frac{\pi}{4}$, $\frac{\pi}{3}$, $\frac{\pi}{2}$).  (b)  Relative entropy of global coherence $C_G$ at $(k_F, k_R) = (10^6, 10^8) s^{-1}$ for two extreme cases $\chi$ ($0$, $\frac{\pi}{2}$) showing an increase in coherence time. The calculations have been done for a six nuclei cryptochrome based radical pair system. }
\label{GCOHRECNE_CISS}
\end{figure}

In Fig.~\ref{GCOHRECNE_CISS}, we have plotted the relative entropy of global coherence $C_G(\rho)$ with respect to time at  $\theta=0$ and $\phi=0$. In Fig.~\ref{GCOHRECNE_CISS}.(a), we have considered five distinct value of $\chi$ showing varying degree of spin selectivity due to CISS ($0$, $\frac{\pi}{6}$, $\frac{\pi}{4}$, $\frac{\pi}{3}$, $\frac{\pi}{2}$). For analysis, we have considered a realistic rate combination $(k_F, k_R) = (10^6, 10^8) s^{-1}$ for 6-nuclei cryptochrome based molecule. In Fig.~\ref{GCOHRECNE_CISS}.(b) we have considered two extreme cases $\chi = $ ($0$,$\frac{\pi}{2}$) highlighting the increase in global coherence (magnitude and time duration).  A  finite value of $C_G$ for a longer duration of time was observed when $\chi=\frac{\pi}{2}$ compared to $\chi=0$.

\begin{figure}[htbp]
\centering
\includegraphics[width=90mm,keepaspectratio]{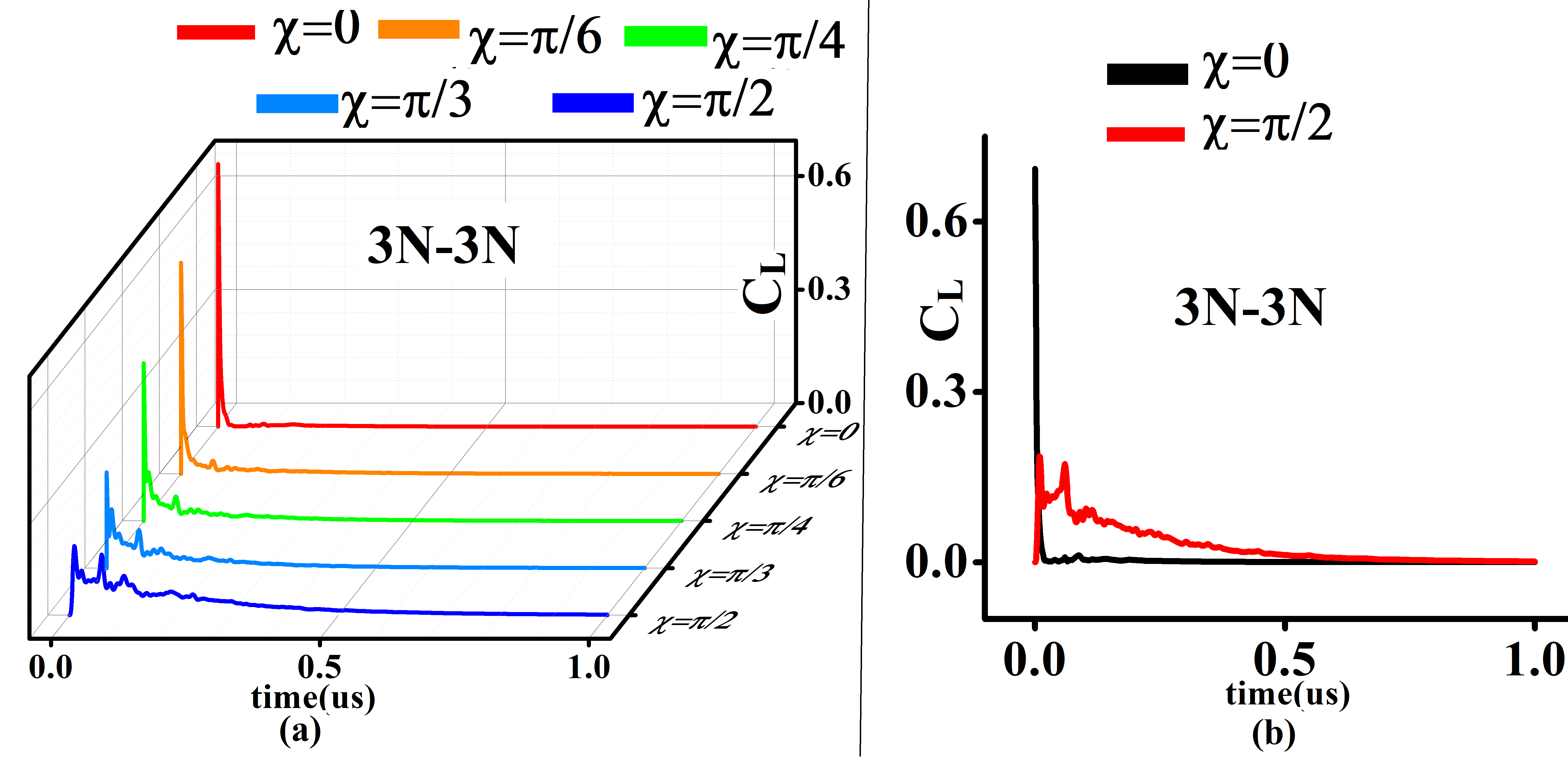}
\caption{(a) Relative entropy of local coherence $C_L$ at $(k_F, k_R) = (10^6, 10^8) s^{-1}$ for five distinct value of $\chi$ showing varying degree of spin selectivity due to CISS ($0$, $\frac{\pi}{6}$, $\frac{\pi}{4}$, $\frac{\pi}{3}$, $\frac{\pi}{2}$). The calclulations have been done for six-nuclei cryptochrome based radical pair system.  (b) Relative entropy of local coherence $C_L$ at $(k_F, k_R) = (10^6, 10^8) s^{-1}$ for two extreme cases i.e. $\chi = $ ($0$,$\frac{\pi}{2}$) exhibiting sustained coherence.}
\label{LCOHRECNE_CISS}
\end{figure}

In Fig.~\ref{LCOHRECNE_CISS}, we have plotted the relative entropy of local coherence $C_L(\rho)$ with respect to time. In Fig.~\ref{LCOHRECNE_CISS}.(a), we have considered five distinct value of $\chi = (0$, $\frac{\pi}{6}$, $\frac{\pi}{4}$, $\frac{\pi}{3}$, $\frac{\pi}{2}$) for a  realistic rate combination of $(k_F, k_R) = (10^6, 10^8) s^{-1}$ for 6-nuclei cryptochrome molecule based radical pair system. In Fig.~\ref{LCOHRECNE_CISS}.~(b), we have considered two extreme cases i.e. $\chi = $ ($0$,$\frac{\pi}{2}$) highlighting the increment in local coherence with CISS. %The $C_L$ quantifier gives us information only about the radical pair electronic coherence whereas $C_G$ is associated with the coherence of the radical pair and surrounding nuclei.
At $t=0$, we observe the maximum value of $C_L$ at $\chi=0$. As the spin selectivity increases, $C_L$ decreases at $t=0$. It can be attributed to the initial value of $\rho$ at $t=0$ i.e. as $\chi$ increases, the non-diagonal terms of density matrix ($\rho$) associated with radical pair local coherence decrease. However, even though at $t=0$, the system is showing maximum local coherence ($C_L$) at $\chi=0$, the case of full CISS $\chi=\frac{\pi}{2}$ shows us the sustained coherence over evolution. Hence, we deduce that the radical pair system has sustained coherence due to CISS.

% =======
% FIG. 0
% =======

\begin{figure}[htbp]
\centering
\includegraphics[width=90mm,keepaspectratio]{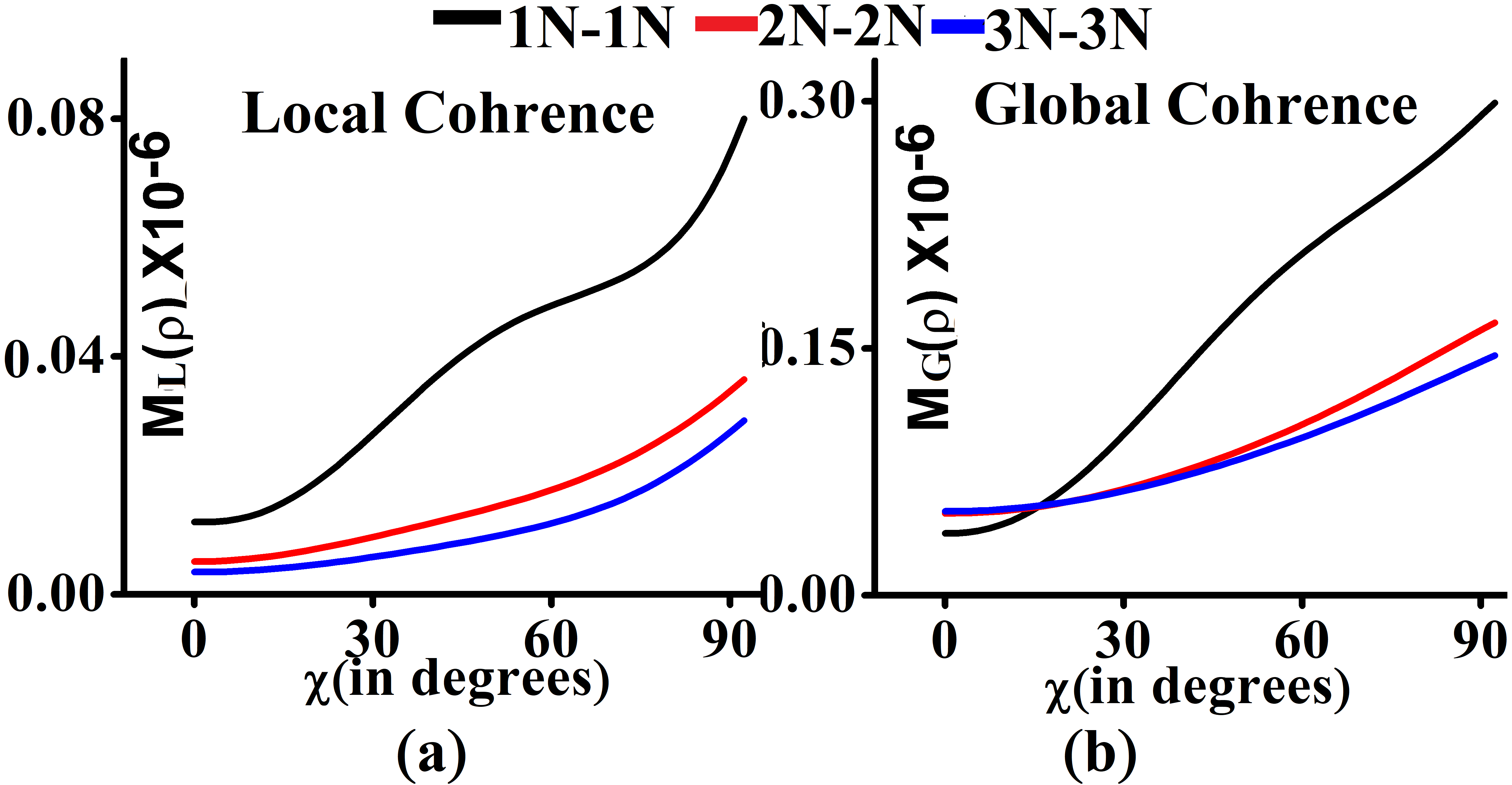}
\caption{Relative Entropy of (a) local coherence ($M_L$) and (b) global cohrence ($M_G$) at $(k_F, k_R) = (10^6, 10^8)~s^{-1}$ for $\chi\in [0,\frac{\pi}{2}]$. This has been done for 2-nuclei (Black), 4-nuclei (Red), and 6-nuclei (Blue) cryptochrome based radical pair system.}
\label{TOTALCOHRECNE_CISS}
\end{figure}

In our investigation from Fig.~\ref{GCOHRECNE_CISS} and Fig.~\ref{LCOHRECNE_CISS}, it was observed that an increase in CISS results in an increase in coherence. We further confirm this by making use of Eq.~\ref{TOTALCOHRENCE} that captures the coherence over entire duration of the spin evolution. We plot $M_{i}$ as function of spin selectivity $\chi$ in Fig.~\ref{TOTALCOHRECNE_CISS} at realistic rates of $(k_F, k_R) = (10^6, 10^8)~s^{-1}$ for 2-nuclei (black), 4-nuclei (red), and 6-nuclei (blue) cryptochrome based radical pair system. 
 
 % =======
% FIG. 0
% =======

As is clear from Fig.~\ref{TOTALCOHRECNE_CISS}, the total coherence $M_i$ ($i \in\lbrace L, G\rbrace$) increases with the degree of CISS. As expected, with the inclusion of more nuclei total coherence of the system decreases for a fixed value of $\chi$. To better analyze this, we define another quantity called $\triangle M_i$ in Eq.\ref{DELTA_MC} to quantify the change in total coherence $M_i$ due to CISS.
 % =======
% EQ. 3
% =======
\begin{equation} 
\label{DELTA_MC}
\triangle M_{i}=\frac{ \max_{\chi\in\lbrace 0^o, 90^o\rbrace}(M_{i})}{ \min_{\chi\in\lbrace 0^o, 90^o\rbrace}(M_{i})}
\end{equation}
Table \ref{table2} gives the value of $\triangle M_{i}$ for all three systems.
We observe that all values are greater than unity, signifying an increase in coherence in all systems. However, the value of $\triangle M_{G}$  decreases as the number of nuclei increases. The above analysis though performed for $\theta=0$ and $\phi=0$, the increase in coherence due to CISS was observed at all orientation of radical with respect to earth's magnetic field.

\begin{table}%
\caption{\label{table2}%
 $\triangle M_i$ for radical pair model based on 2, 4, and 6 nuclei from cryptochrome based radical pair system for rates  $(k_F, k_R) = (10^6, 10^8) s^{-1}$.}
\begin{ruledtabular}
\begin{tabular}{lcdr}
\textrm{Nuclei System}&
\textrm{$\triangle M_{G}$}&
\textrm{$\triangle M_{L}$}\\
\colrule
1N-1N (2-nuclei) & 7.97 & 6.59 \\
2N-2N (4-nuclei) & 3.33 & 6.57 \\
3N-3N (6-nuclei) & 2.86 & %5.81
7.82\\
\end{tabular}
\end{ruledtabular}
\end{table}

\subsection{\label{CISSCOHRENCEEE}Effect of Dipolar Interaction}

This subsection studies the effect of spin dipolar interaction along with CISS on coherence in the radical pair system. %Both of these effects arise due to the spin property of the electrons.
Dipolar interaction ($D$) is governed by Eq.~\ref{Dipolar_Interaction}  where $r$ is the distance between two electrons~\cite{efimova2008role}.
% =======
% EQ. 4
% =======
\begin{equation} 
\label{Dipolar_Interaction}
\begin{split}
D(r)=-\frac{3}{2}\frac{\mu_o}{4\pi}\frac{\gamma^2_e\hbar^2}{r^3} \Rightarrow
D(r)/\mu T=-\frac{2.78 \times 10^3}{(r/nm)^3}
\end{split}
\end{equation}

\begin{figure}[htbp]
\centering
\includegraphics[width=90mm,keepaspectratio]{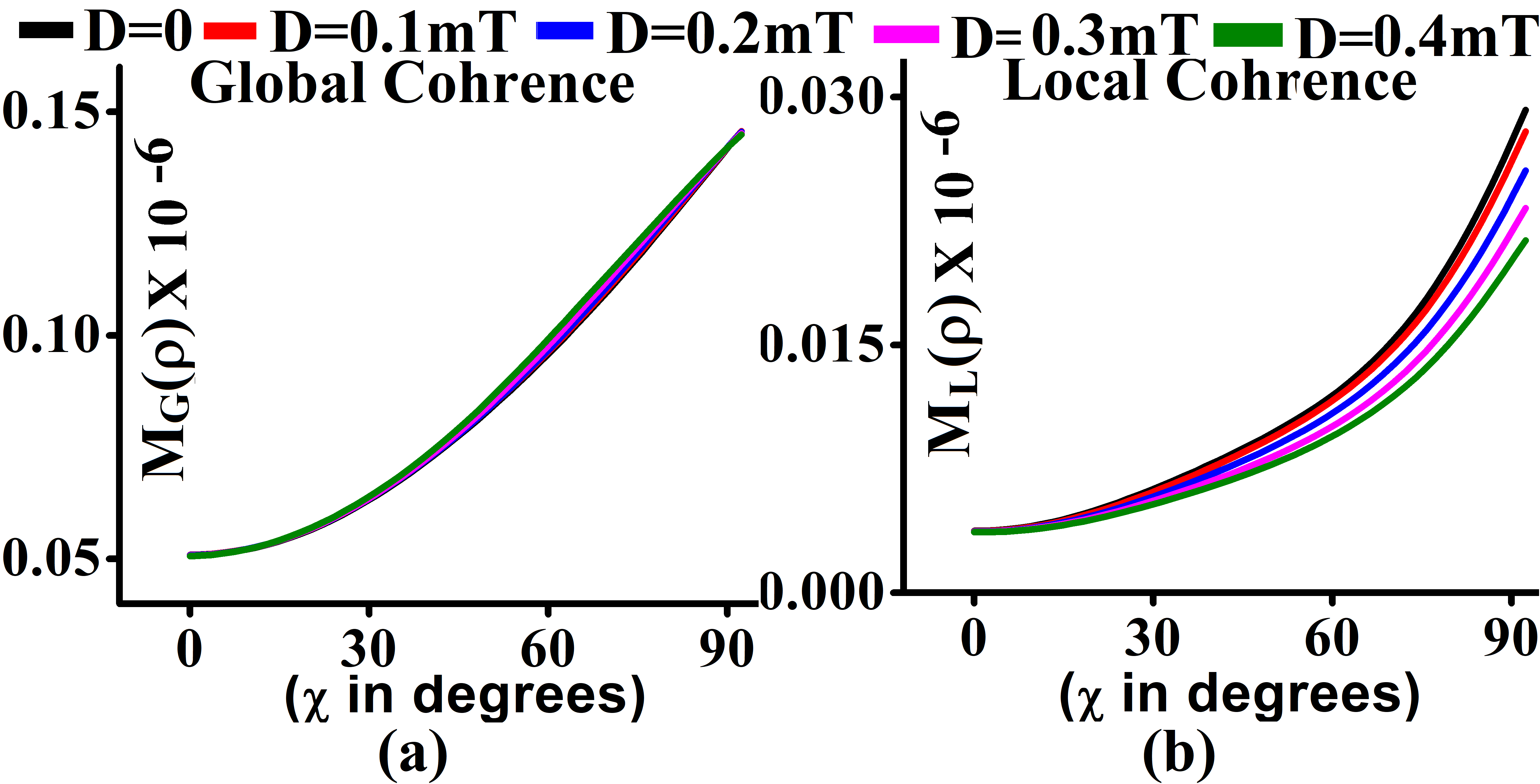}
\caption{Relative entropy of (a) global coherence and (b) local coherence at $(k_F, k_R) = (10^6, 10^8)~s^{-1}$ for $\chi\in [0,\frac{\pi}{2}]$. A total of five values of $D$ were assumed (0, 0.1mT, 0.2mT, 0.3mT, and 0.4mT). The exchange interaction was assumed negligible here. The calculations have been done for six nuclei from cryptochrome molecule at $\theta=0$ and $\phi=0$.}
\label{COHRECNE_CISS}
\end{figure}

We plot Fig.~\ref{COHRECNE_CISS} to study the effect of dipolar interaction on total global and local coherence measures ($M_{G}$ and $M_{L}$). We plot $M_{G}$ (Fig.~\ref{COHRECNE_CISS}.~a) and $M_{L}$ (Fig.\ref{COHRECNE_CISS}.b) with respect to $\chi$. We plot for five distinct values of dipolar interaction, assuming there is no exchange interaction for the six nuclei from cryptochrome molecule. We take the realistic set of rates ($(k_F, k_R) = (10^6, 10^8)~s^{-1}$) in our analysis.
% =======
% FIG. 0
% =======

From both plots ibn Fig.\ref{COHRECNE_CISS}, we observe that the increase in global coherence due to CISS ($\triangle M_{G}$) remains constant and is unaffected due to dipolar interaction. However, $\triangle M_{L}$ is affected by dipolar interaction and decreases about 13\% as $D$ increases from 0 to 0.4mT. It is summarized and confirmed in Table~\ref{table1}. In Fig.~\ref{COHRECNE_CISS}.~(a), we observe that for intermediate values of $\chi$, having dipolar interaction increases the total global coherence. To analyze this further, we need to define a  quantity called $\triangle G_{D=i}(\chi)$ and $\triangle L_{D=i}(\chi)$ given in Eq.~\ref{DeltaG} and Eq.~\ref{DeltaL} respectively. $\triangle G_{D=i}(\chi)$ and ($\triangle L_{D=i}(\chi)$) compute the difference of total global (local) coherence when $D=0$ and when $D=i$ where $i\in\lbrace$ 0.1mT, 0.2mT,0.3mT,0.4mT$\rbrace$ at a particular $\chi$.

\begin{table}%
\caption{\label{table1}%
 $\triangle M_{i}$ for radical pair model based on six nuclei from cryptochrome molecule for rates $(k_F, k_R) = (10^6, 10^8) s^{-1}$.}
\begin{ruledtabular}
\begin{tabular}{lcdr}
\textrm{Dipolar Interaction}&
\textrm{$\triangle M_{G}$}&
\textrm{$\triangle M_{L}$}\\
\colrule
D=0 & 2.86 & 7.82 \\
D=0.1mT & 2.86 & 7.50 \\
D=0.2mT & 2.86 & 6.90 \\
D=0.3mT & 2.86 & 6.31 \\
D=0.4mT & 2.86 & 5.81 \\
\end{tabular}
\end{ruledtabular}
\end{table}

% =======
% EQ. 4
% =======
\begin{equation} 
\label{DeltaG}
\begin{split}
\triangle G_{D=i}(\chi) = M_{G,D=0}(\chi) - M_{G,D=i}(\chi) 
\end{split}
\end{equation}
% =======
% EQ. 4
% =======
\begin{equation} 
\label{DeltaL}
\begin{split}
\triangle L_{D=i}(\chi) = M_{L,D=0}(\chi) - M_{L,D=i}(\chi) 
\end{split}
\end{equation}

\begin{figure}[htbp]
\centering
\includegraphics[width=90mm,keepaspectratio]{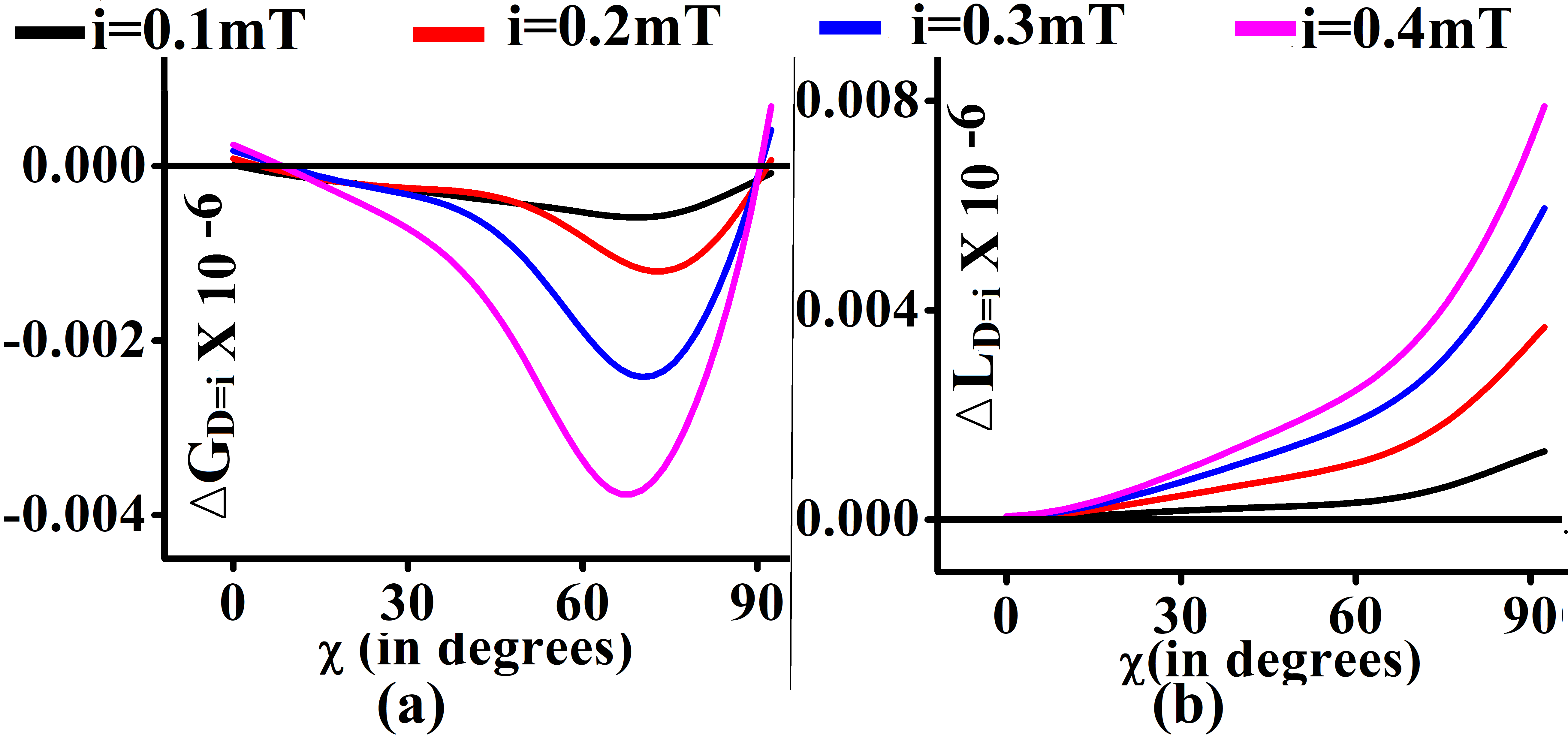}
\caption{(a)$\triangle G_{D=i}$ (b)$\triangle L_{D=i}$ where $i$ is 0.1mT (black), 0.2mT (red), 0.3mT (blue), 0.4mT (pink) at $(k_F, k_R) = (10^6, 10^8) s^{-1}$ for $\chi\in [0,\frac{\pi}{2}]$. The horizontal dotted line is the reference line depicting when $\triangle G_{D=i}=0$ and $\triangle L_{D=i}=0$. Anything above this line shows that total coherence is greater for the case when $D = 0$ than when $D=i$. This has been done for six nuclei from cryptochrome molecules at $\theta=0$ and $\phi=0$.}
\label{COHRECNE_CISSGAP}
\end{figure}

In Fig.~\ref{COHRECNE_CISSGAP}, we plot $\triangle G_{D=i}$ and $\triangle L_{D=i}$ as function of $\chi$. A horizontal reference line in Fig.~\ref{COHRECNE_CISSGAP} depicts $\triangle G_{D=i}=0$ and $\triangle L_{D=i}=0$. Anything above this line shows that total coherence is greater when $D = 0$ than $D=i$. Fig.~\ref{COHRECNE_CISSGAP}.~(a) plots $\triangle G_{D=i}$  where for intermediate values of $\chi$ we observe a negative value of $\triangle G_{D=i}$. It signifies that dipolar interaction enhances global coherence for these values of $\chi$. The range of values of $\chi$ for which we observe an increase in coherence is approximately the same for all values of $D$. 
Fig.~\ref{COHRECNE_CISSGAP}.(b) discusses $\triangle L_{D=i}$ where we observe that  $\triangle L_{D=i}$ is always positive for all values of $D$. Hence for all values of dipolar interactions, local coherence shows degradation in total local coherence $M_L$. The  exchange interaction ($J$) further increases total global coherence which is discussed in detail in Appendix~\ref{AppendixA}.

% =======
% FIG. 0
% =======

\subsection{\label{DIFFRENT_RATE_COHRENCE}Coherence in RP System for Various Rates}
In this subsection, we ascertain the increase in coherence with CISS at different rates. We present Tab~\ref{table3G} and Tab~\ref{table3l} that show an increase in total coherence due to CISS (through $\triangle M_G$ and $\triangle M_L$) for a wide range of rate combinations. We found that maximum value of $\triangle M_G$ is at $(k_F, k_R) = (10^4, 10^8)~s^{-1}$ and $\triangle M_L$ is at $(k_F, k_R) = (10^4, 10^6)~s^{-1}$ (maxima occurs at a different rate combination). However, interestingly a lower $k_F$ (protonation rate) is key to achieve maximum increase in coherence due to CISS.   

\begin{table}%
\caption{\label{table3G}%
 $\triangle M_{G}$ for global coherence for radical pair model based on six nuclei from cryptochrome molecule for various rate combination at $D=0$ and $J=0$.
}
\begin{ruledtabular}
\begin{tabular}{llllll}
\textrm{$k_R\downarrow, k_F\rightarrow$}&
\textrm{$10^4 s^{-1}$}&
\textrm{$10^5 s^{-1}$}&
\textrm{$10^6 s^{-1}$}&
\textrm{$10^7 s^{-1}$}&
\textrm{$10^8 s^{-1}$}\\
\colrule
$10^4 s^{-1}$ & 0.98 & 0.87 & 0.86 & 0.84 & 0.68 \\
$10^5 s^{-1}$ & 1.57 & 0.98 & 0.86 & 0.84& 0.68 \\
$10^6 s^{-1}$ & 2.44 & 1.57 & 0.98 & 0.86& 0.68 \\
$10^7 s^{-1}$ & 2.78 & 2.50 & 1.61 & 0.99 &0.71 \\
$10^8 s^{-1}$ & 3.82 & 3.74 & 2.86 & 2.01& 0.98 \\

\end{tabular}
\end{ruledtabular}
\end{table}

\begin{table}%
\caption{\label{table3l}%
 $\triangle M_{L}$ for local coherence for radical pair model based on six nuclei from cryptochrome molecule for various rate combination at $D=0$ and $J=0$.
}
\begin{ruledtabular}
\begin{tabular}{llllll}
\textrm{$k_R\downarrow, k_F\rightarrow$}&
\textrm{$10^4 s^{-1}$}&
\textrm{$10^5 s^{-1}$}&
\textrm{$10^6 s^{-1}$}&
\textrm{$10^7 s^{-1}$}&
\textrm{$10^8 s^{-1}$}\\
\colrule
$10^4 s^{-1}$ & 4.53 & 2.83 & 2.52 &1.51& 0.33 \\
$10^5 s^{-1}$ & 13.76 & 4.52 & 2.66 & 1.52& 0.33 \\
$10^6 s^{-1}$ & 19.49 & 13.33 & 4.13 & 1.58& 0.33 \\
$10^7 s^{-1}$ & 18.56 & 16.98 & 10.04 & 2.10 &0.33 \\
$10^8 s^{-1}$ & 9.10 & 8.82 & %7.58%
7.82 & 2.88& 0.37 \\

\end{tabular}
\end{ruledtabular}
\end{table}
\section{\label{Environmental_Decoherence} Effect of Environmental Decoherence}

In this section, we take into consideration the decoherence effect of the surrounding system. We modify Eq.~\ref{MASTER_EQUATION} to add spin decoherence operators in the Lindblad formalism.
% =======
% EQ. 4
% =======
\begin{equation} 
\label{MASTER_EQUATIONN}
\begin{split}
{\frac{d\hat{\rho}}{dt}=-(Coherent  + Recombination. + Decohrence)}\\
=-i[\hat{H},\hat{\rho}(t)]-\frac{1}{2}k_R[P_R,\hat{\rho}(t)]-k_F\hat{\rho}(t)\\
+k\sum_{n}\frac{1}{2}\{ 2C_{n}\rho(t)C_{n}^\dagger-\rho(t)C_{n}^\dagger C_{n}-C_{n}^\dagger C_{n}\rho(t)\}
\end{split}
\end{equation}

In Eq.~\ref{MASTER_EQUATIONN}, $Decohrence$ corresponds to the spin decoherence occurring due to surrounding environment. Mathematically we take six decoherence operators: $C_1 = \sigma_x \otimes I_{E2} \otimes I_{N}$,  $C_2 = \sigma_y \otimes I_{E2} \otimes I_{N}$, $C_3 = \sigma_z \otimes I_{E2} \otimes I_{N}$, $C_4 = I_{E1} \otimes \sigma_x \otimes I_{N}$,$C_5 = I_{E1} \otimes \sigma_y \otimes I_{N}$ and $C_6 = I_{E1} \otimes \sigma_z \otimes I_{N}$. $I_{E1}$ correspond to the mixed state of electron on FAD$^\cdot$ radical while $I_{E2}$ correspond to the mixed state of electron on TrpH$^\cdot$ radical. $I_{N}$ is the combined mixed state of the nuclei and $k$ is the decoherence rate.
% =======
% FIG. 0
% =======
\begin{figure}[htbp]
\centering
\includegraphics[width=90mm,keepaspectratio]{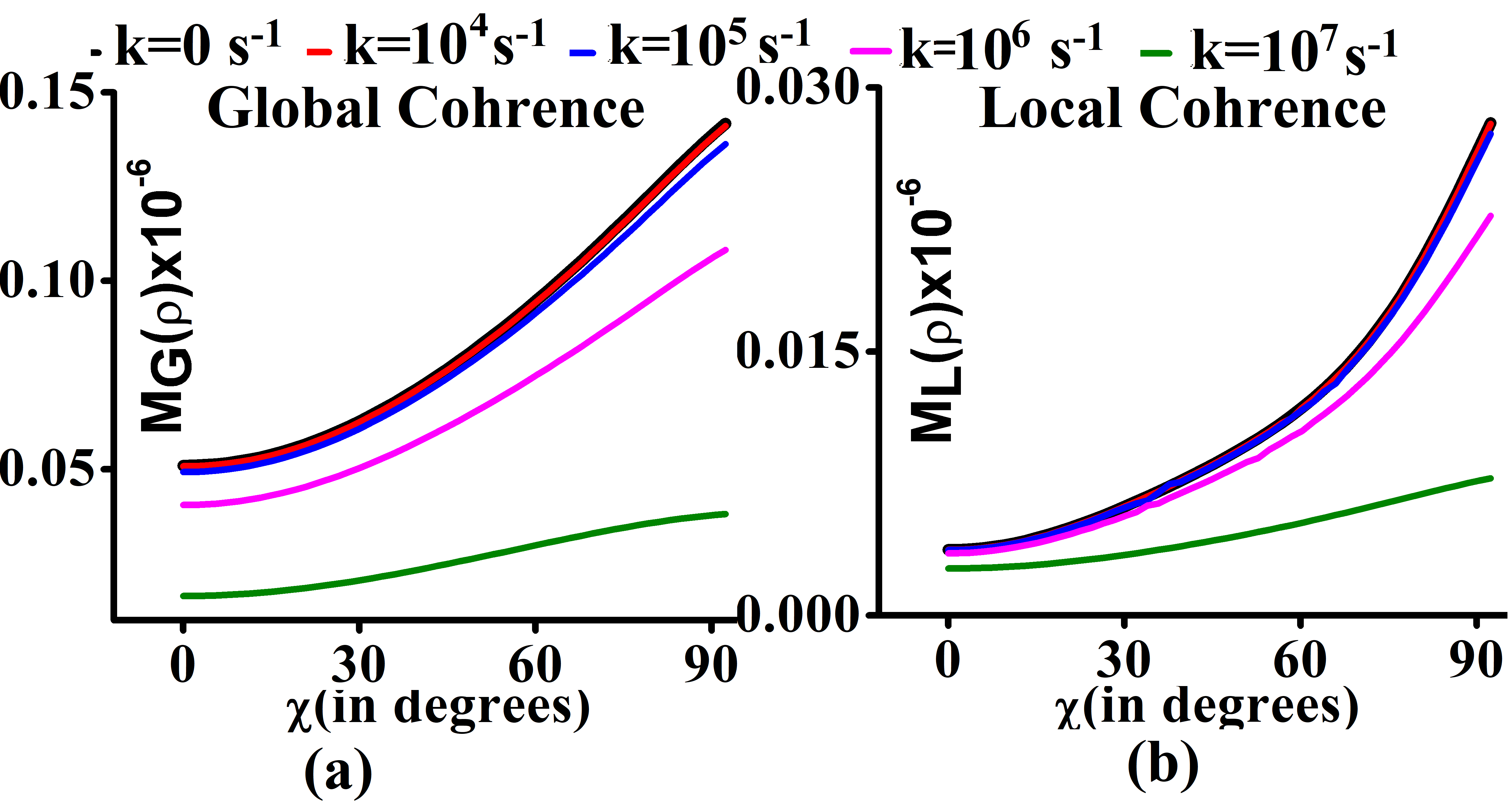}
\caption{(a) $M_{G}(\chi)$, (b) $M_{L}(\chi)$ where k is the decoherence rate $k=0~s^{-1}$ (black), $k=10^4~s^{-1}$ (red), $k=10^5~s^{-1}$ (blue), $k=10^6~s^{-1}$ (pink) and $k=10^7~s^{-1}$ (green). The calculations have been done for six-nuclei cryptochrome based RP system. We have assumed $D=0$ and $J=0$ here.}
\label{COHRECNE_RELAXATION}
\end{figure}

\begin{table}%
\caption{\label{table3}%
 $\triangle M_{i}$ for radical pair model based on six nuclei from cryptochrome molecule for various relaxation rate at $D=0$ and $J=0$  $(k_F, k_R) = (10^6, 10^8) s^{-1}$  
}
\begin{ruledtabular}
\begin{tabular}{lcdr}
\textrm{Relaxation rate k}&
\textrm{$\triangle M_{G}$}&
\textrm{$\triangle M_{L}$}\\
\colrule
$k=0s^{-1}$ & 2.86 & 7.82 \\
$k=10^4s^{-1}$ & 2.78 & 7.50 \\
$k=10^5s^{-1}$ & 2.76 & 7.40 \\
$k=10^6s^{-1}$ & 2.67 & 6.44 \\
$k=10^7s^{-1}$ & 2.32 & 2.93 \\

\end{tabular}
\end{ruledtabular}
\end{table}

In Fig.~\ref{COHRECNE_RELAXATION}, we plot total global ($M_{G}$) and local cohrence ($M_{L}$). This calculation has been done at $(k_F, k_R) = (10^6, 10^8)~s^{-1}$ and $J=0$ and $D=0$. We observe that at full CISS, coherence is maximum even under decoherence.  We have also listed $\triangle M_{G}$ and $\triangle M_{L}$ with various decoherence rates $k$ in Tab.~\ref{table3}. We observe a reduction in value of $\triangle M_{G}$ and $\triangle M_{L}$ as decoherence rate $k$ increases. However, interestingly, we observe increment in coherence due to CISS even at high decoherence rates.

\section{\label{Utilitarian}On Utilitarian role of coherence}
In this section, we correlate coherence to the forward product (signalling state) yield of the reaction described in Fig.~\ref{RECOMBINATION}, demonstrating the utilitarian role of coherence.  We use the correlation coefficient to show a statistical correlation between signaling state yield and the radical pair spin coherence (both local and global). We use numerous orientations of radical pair with respect to the external magnetic field to show this correlation in a four nuclei system at $(k_F, k_R) = (10^6, 10^8)~s^{-1}$. The forward (signaling state) yield is defined as:
% =======
% EQ. 4
% =======
\begin{equation} 
\label{YIELD_PARTIALSTATE}
\begin{split}
\phi_F=k_F\int_{0}^{\infty} P_S(t)dt=k_F\int_{0}^{\infty} Tr[\hat{\rho(t)}]dt
\end{split}
\end{equation}
%====================
%The forward signaling state is responsible for the signal sent to the brain (see Fig.~\ref{RECOMBINATION}).
%The yield of the signaling state ($\phi_F$) is defined by Eq.~\ref{YIELD_PARTIALSTATE} 
Where $\hat{\rho(t)}$ is the solution of the master equation Eq.~\ref{MASTER_EQUATION}, $Tr$ is the trace over the state density matrix $\rho$. $k_F$ is the rate associated with the signaling state. 

% =======
% FIG. 0
% =======
\begin{figure}[htbp]
\centering
\includegraphics[width=90mm,keepaspectratio]{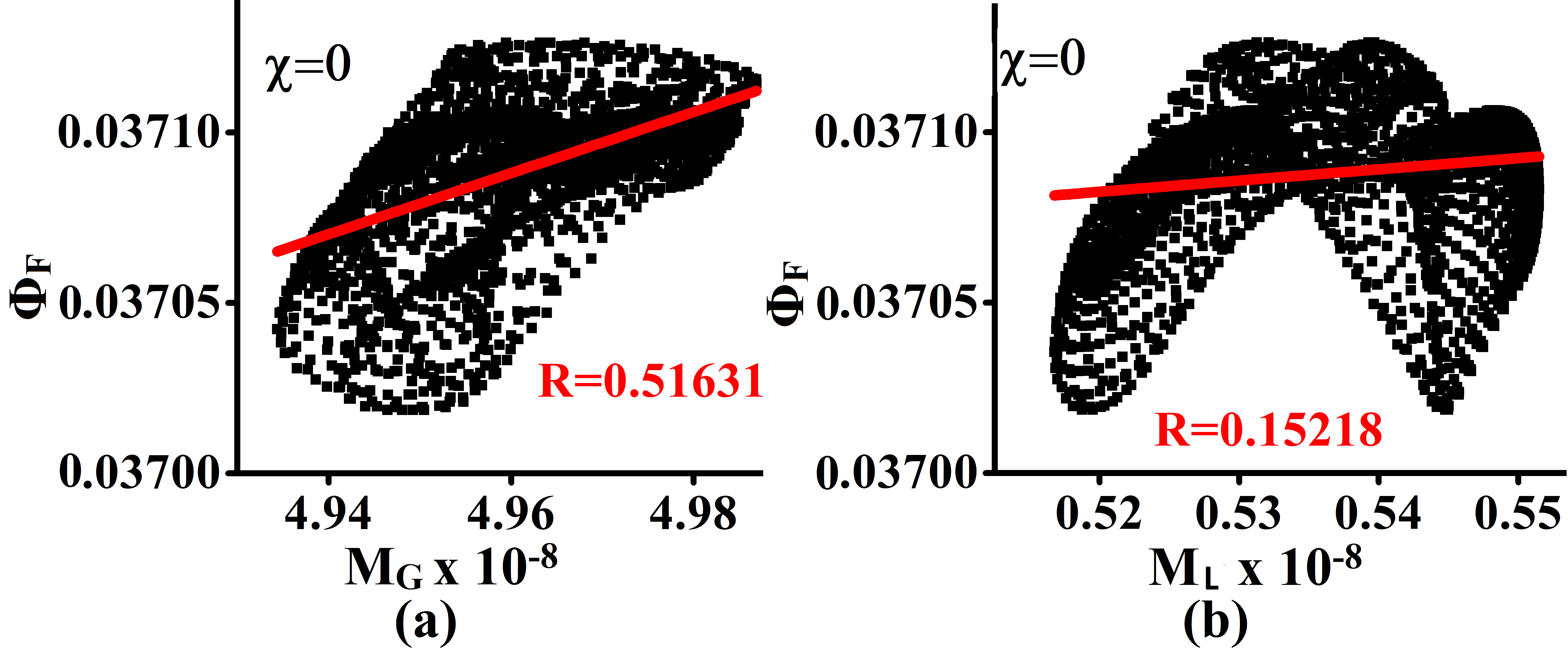}
\caption{(a) $M_{G}$ vs $\phi_F$, and (b) $M_{L}$ vs $\phi_F$ for various values of $\theta$ and $\phi$ for $\chi=0$. The calculation is done for four-nuclei from cryptochrome based RP system at $(k_F, k_R) = (10^6, 10^8)~s^{-1}$ with $D=0$ and $J=0$. The red line corresponds to the linear fit, and $R$ (red) corresponds to the correlation coefficient between the coherence measure and signaling state yield. }
\label{Correlationu0}
\end{figure}

\begin{figure}[htbp]
\centering
\includegraphics[width=90mm,keepaspectratio]{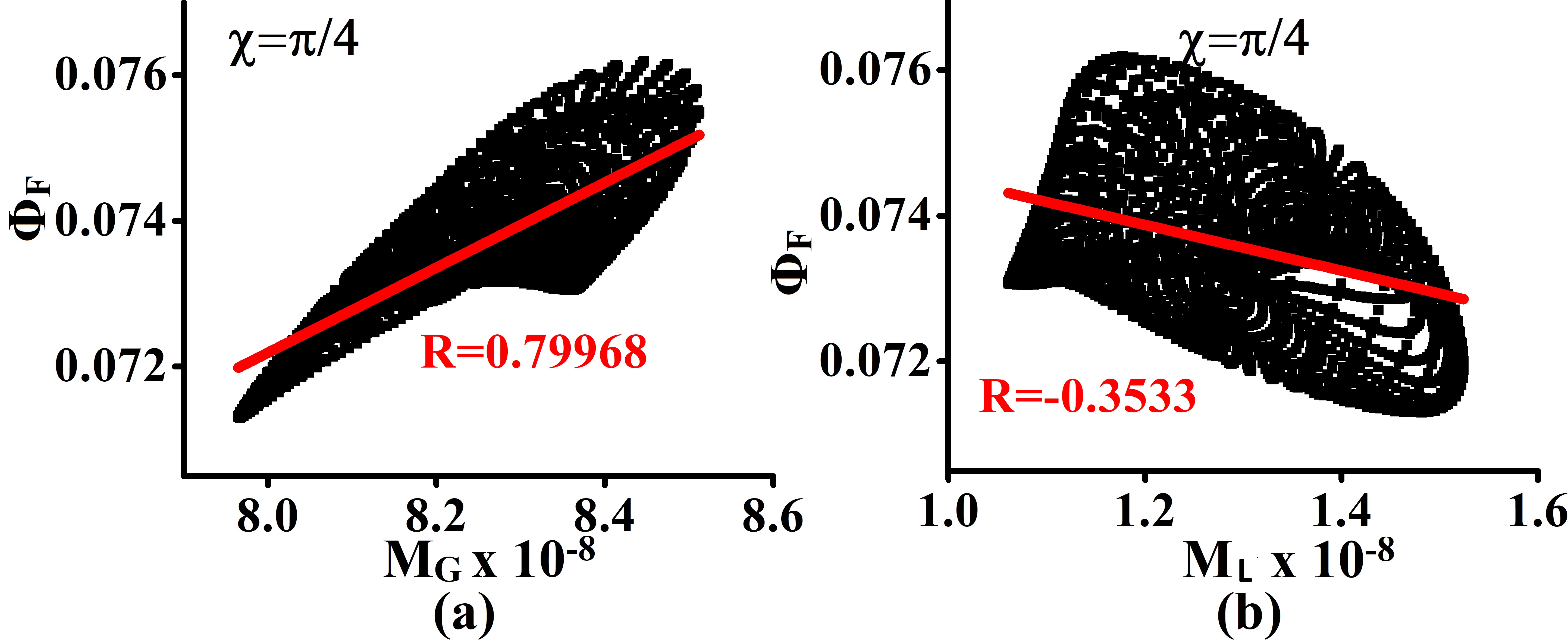}
\caption{(a) $M_{G}$ vs $\phi_F$, and (b) $M_{L}$ vs $\phi_F$ for various values of $\theta$ and $\phi$ for $\chi=\frac{\pi}{4}$. The calculation is performed for four-nuclei cryptochrome based RP system at $(k_F, k_R) = (10^6, 10^8)~s^{-1}$ with $D=0$ and $J=0$. The red line corresponds to the linear fit, and $R$ (red) corresponds to the correlation coefficient between the coherence measure and signaling state yield. }
\label{Correlationupi4}
\end{figure}

\begin{figure}[htbp]
\centering
\includegraphics[width=90mm,keepaspectratio]{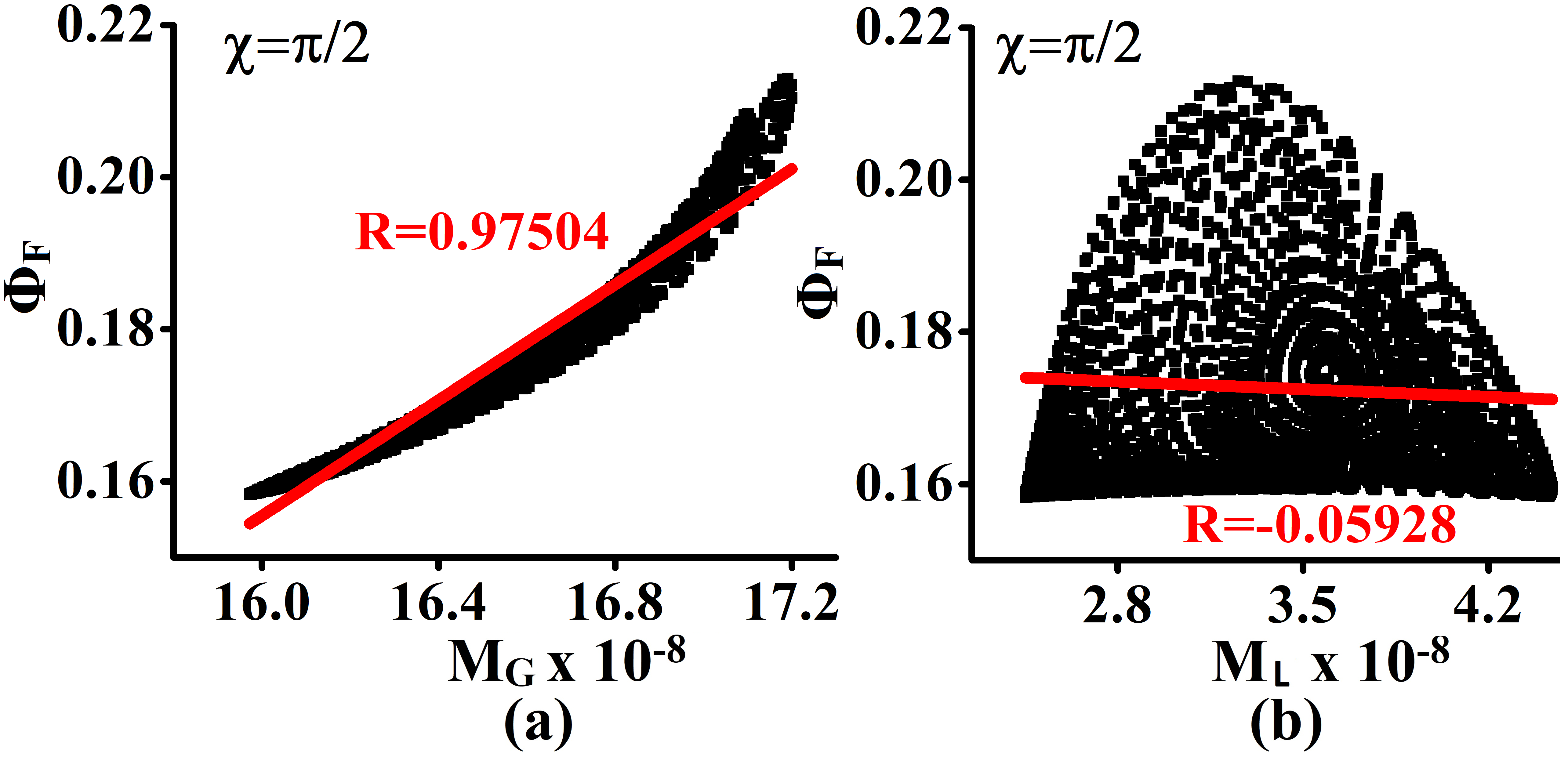}
\caption{(a)$M_{G}$ vs $\phi_F$ and (b)$M_{L}$ vs $\phi_F$ for various values of $\theta$ and $\phi$ for $\chi=\frac{\pi}{2}$. The calculation is performed for four-nuclei from cryptochrome based RP system at $(k_F, k_R) = (10^6, 10^8) s^{-1}$ with $D=0$ and $J=0$. The red line corresponds to the linear fit, and $R$ (red) corresponds to the correlation coefficient between the coherence measure and signaling state yield.}
\label{Correlationupi2}
\end{figure}

In Fig.~\ref{Correlationu0}.(a), we have plotted the total global coherence ($M_G$) and the signaling state yield ($\phi_F$) for 2500 combinations of $\theta$ and $\phi$. We have taken the values where $\theta\in\lbrace 0^o, 180^o\rbrace$ and $\phi\in\lbrace 0^o, 360^o\rbrace$. The calculation is performed for no CISS case i.e. $\chi=0$. Similarly, in Fig.~\ref{Correlationu0}.(b), we have plotted total local coherence ($M_L$) and signaling state yield ($\phi_F$). The $R$ value (red font) corresponds to the correlation coefficient between the coherence measure and signaling state yield. The red line is the linear fit line corresponding to the scattered points. Similar plots have been plotted for intermediate CISS case (i.e. $\chi=\frac{\pi}{4}$) in Fig.~\ref{Correlationupi4} and full CISS case (i.e. $\chi=\frac{\pi}{2}$) in Fig.~\ref{Correlationupi2}.

% =======
% FIG. 0
% =======

% =======
% FIG. 0
% =======

The total local coherence ($M_L$) has no clear correlation with the signaling state yield ($\phi_F$) for three values of $\chi$. %it is not a correct measure to describe the action of the CISS-assisted avian compass. 
Total global coherence ($M_G$) shows a high correlation with the yield of the forward signaling state i.e. as the degree of CISS increases, the correlation parameter $R$ between total global coherence and signaling state yield increases. In the full CISS case, the value is near unity showing a high correlation of global coherence with the forward signaling state. This forward signaling state is responsible for sending signals to neurons or brains. Hence, global coherence exhibit strong utilitarian role for the compass action than the local coherence i.e. a higher spin selectivity due to chirality leads to a higher correlation of global coherence with forward signaling state yield. %In a nutshell, a higher CISS leads to higher global coherence. And due to the high correlation with the yield of the signaling state, we obtain higher sensitivity for the compass.

\section{Conclusion}
In conclusion, chiral-induced spin selectivity (CISS) effect causes sustained coherence in the radical pair mechanism of avian magnetoreception. It hints towards the possibility that spin coherence might be sustained in a realistic system despite many nuclei for significant time. Moreover, we also observe that the global coherence in the CISS-assisted avian compass is strongly correlated with the signaling state yield. This indicates that unlike local coherence. global coherence has strong utilitarian role in the compass action.  We also observe that dipolar and exchange interactions are generally detrimental to the coherence of the avian compass, but their effect can be countered by CISS. All these conclusions confirm the significance of CISS in the avian compass spin dynamics.

\begin{acknowledgments}
%The author would like to thank Jiate Luo, Department of Chemistry, University of Oxford, UK and Peter J. Hore, Department of Chemistry, University of Oxford, UK for their invaluable and insightful communication. 
This work is supported by the Science and Engineering Research Board, Department of Science and Technology (DST), India with grant No. CRG/2021/007060 and DST/INSPIRE/04/2018/000023. The authors would also like to thank Department of Electronics and Communication, IIT Roorkee and Ministry of Education, Government of India for supporting Y.T.'s graduate research.

\end{acknowledgments}
\appendix
\section{Effect of Exchange Interaction}
\label{AppendixA}

In order to understand the role of exchange interaction in six nuclei-based cryptochrome systems, we assume a fixed value of dipolar interaction of $D=0.4mT$. We consider four values of exchange interaction which are less than dipolar interaction ($J=0, J=0.1mT, J=0.2mT, J=0.3mT$) \cite{efimova2008role}.

% =======
% FIG. 0
% =======
\begin{figure}[htbp]
\centering
\includegraphics[width=90mm,keepaspectratio]{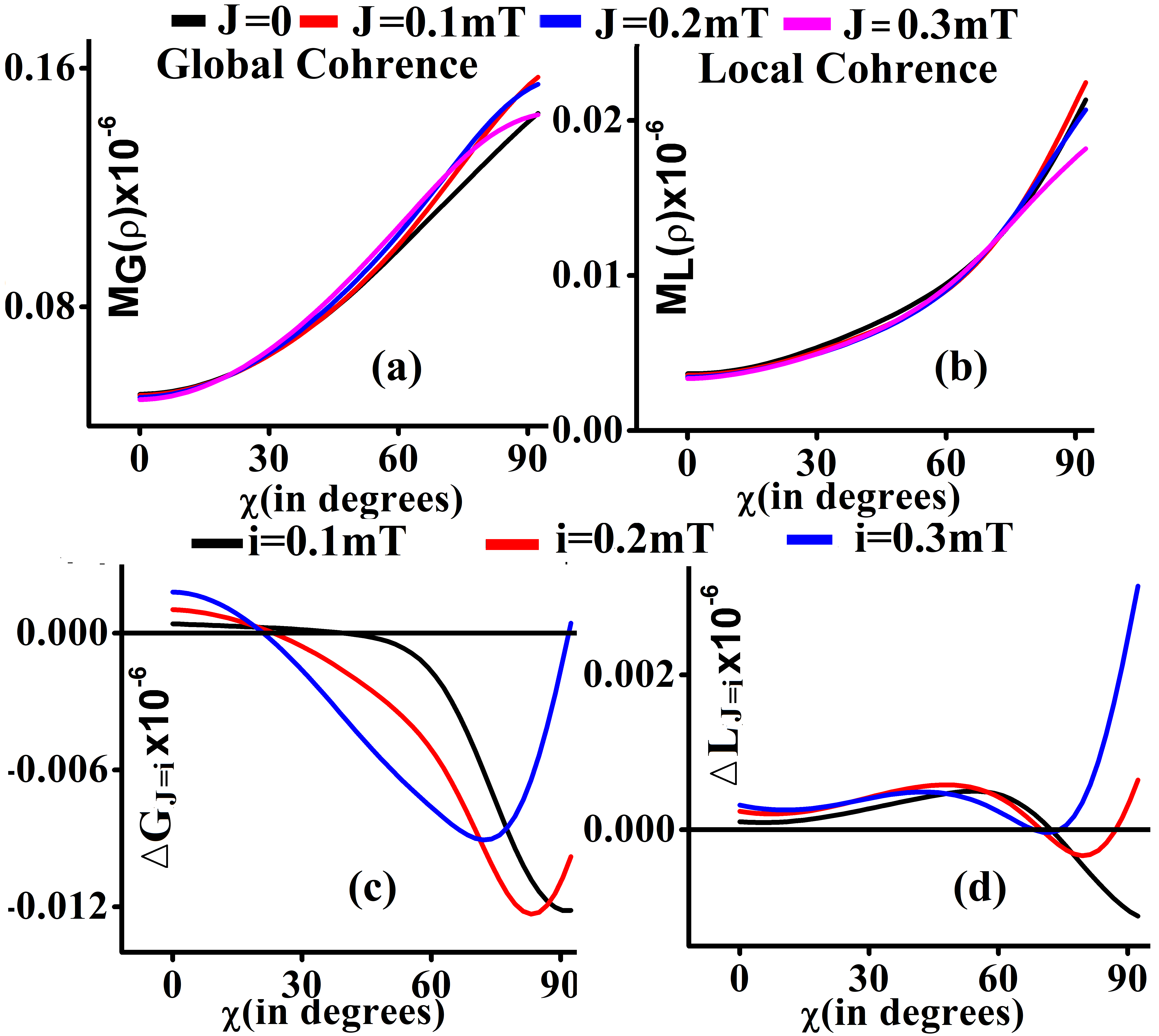}
\caption{Relative Entropy of (a) global Coherence  and (b) local Coherence  at $(k_F, k_R) = (10^6, 10^8) s^{-1}$ for $\chi\in [0,\frac{\pi}{2}]$. A total of four values of $J$ was assumed (0, 0.1mT, 0.2mT, 0.3mT). (c)$\triangle G_{J=i}$ (d)$\triangle L_{J=i}$ where $i$ is 0.1mT (black), 0.2mT (red), 0.3mT (blue) at $(k_F, k_R) = (10^6, 10^8) s^{-1}$. The horizontal dotted line is the reference line depicting when $\triangle G_{J=i}=0$ and $\triangle L_{J=i}=0$. A value above this depicts total coherence under no exchange interaction ($J=0$) having a value greater than when $J=i$. The analysis has been done for $D=0.4mT$.  This has been done for six nuclei from a cryptochrome molecule at $\theta=0$ and $\phi=0$. }
\label{COHRECNE_CISSGAPF}
\end{figure}

In Fig.\ref{COHRECNE_CISSGAPF}, we have plotted $M_G$ and $M_L$ for four distinct values of exchange interaction at $D=0.4mT$ for six nuclei from cryptochrome. We take the realistic set of rates ($(k_F, k_R) = (10^6, 10^8) s^{-1}$) in our analysis. Corresponding to Fig.\ref{COHRECNE_CISSGAPF}.(a),(b)  we have summarized in Table \ref{exchange} the increase in total global( $\triangle M_{G}$) and local coherence( $\triangle M_{L}$) due to CISS. We observe that  $\triangle M_{G}$ and $\triangle M_{L}$ first increase and then decreases with exchange interaction. To analyze the effect of exchange interaction for intermediate cases of CISS, we define a quantity as follows. 

\begin{table}%
\caption{\label{exchange}%
 $\triangle M_{i}$ for radical pair model based on 6 nuclei from cryptochrome molecule for rate at $D=0.4mT$,  $(k_F, k_R) = (10^6, 10^8) s^{-1}$.
}
\begin{ruledtabular}
\begin{tabular}{lcdr}
\textrm{Dipolar Interaction}&
\textrm{$\triangle M_{G}$}&
\textrm{$\triangle M_{L}$}\\
\colrule
J=0 & 2.86 & 5.81 \\
J=0.1mT & 3.12 & 6.29 \\
J=0.2mT & 3.11 & 6.03 \\
J=0.3mT & 2.95 & 5.42 \\

\end{tabular}
\end{ruledtabular}
\end{table}

% =======
% EQ. 4
% =======
\begin{equation} 
\label{DeltaGJ}
\begin{split}
\triangle G_{J=i}(\chi) = M_{G,J=0}(\chi) - M_{G,J=i}(\chi) 
\end{split}
\end{equation}
% =======
% EQ. 4
% =======
\begin{equation} 
\label{DeltaLJ}
\begin{split}
\triangle L_{J=i}(\chi) = M_{L,J=0}(\chi) - M_{L,J=i}(\chi) 
\end{split}
\end{equation}

According to Eq.~\ref{DeltaGJ}, $\triangle G_{J=i}(\chi)$ compute the difference of total global coherence when their is $J=0$ and when $J=i$ where  $i\in\lbrace 0.1mT, 0.2mT,0.3mT\rbrace$ at a particular $\chi$. A similar quantity is defined for total local coherence $\triangle L_{J=i}(\chi)$ in Eq.~\ref{DeltaLJ} . In Fig.~\ref{COHRECNE_CISSGAPF}(c),(d).  we plot the $\triangle G_{J=i}(\chi)$ and $\triangle L_{J=i}(\chi)$ for a fixed value of $D=0.4mT$. A reference horizontal line is drawn depicting $\triangle G_{J=i}=0$ and $\triangle L_{J=i}=0$. Anything above this line depicts that total coherence is greater for the case when $J = 0$ compared to when $J$ has finite value $i$.

Fig.~\ref{COHRECNE_CISSGAPF}.(c) discuss the difference in $\triangle G_{J=i}$ where we observe that for intermediate value of $\chi$, we obtain a negative value. Exchange interaction further enhances global coherence for these values of $\chi$. The window of values of $\chi$ for which we observe an increase in coherence increases with exchange being maximum for $J=0.3mT$. The increase in coherence is, however, maximum for case $J=0.1mT$. Fig.~\ref{COHRECNE_CISSGAPF}.(d) discuss the difference in $\triangle L_{J=i}$ where we observe that for certain $\chi$, we get a negative value. The window of values of $\chi$ for which we observe an increase in local coherence decrease with exchange being maximum for $J=0.1mT$. That means for these values of $\chi$, exchange interaction enhances local coherence. Hence we observe that for intermediate cases of CISS, total coherence shows an improvement  when dipolar and exchange interaction affects the radical pair. 
\bibliography{apssamp}% Produces the bibliography via BibTeX.

\begin{thebibliography}{29}%
\makeatletter
\providecommand \@ifxundefined [1]{%
 \@ifx{#1\undefined}
}%
\providecommand \@ifnum [1]{%
 \ifnum #1\expandafter \@firstoftwo
 \else \expandafter \@secondoftwo
 \fi
}%
\providecommand \@ifx [1]{%
 \ifx #1\expandafter \@firstoftwo
 \else \expandafter \@secondoftwo
 \fi
}%
\providecommand \natexlab [1]{#1}%
\providecommand \enquote  [1]{``#1''}%
\providecommand \bibnamefont  [1]{#1}%
\providecommand \bibfnamefont [1]{#1}%
\providecommand \citenamefont [1]{#1}%
\providecommand \href@noop [0]{\@secondoftwo}%
\providecommand \href [0]{\begingroup \@sanitize@url \@href}%
\providecommand \@href[1]{\@@startlink{#1}\@@href}%
\providecommand \@@href[1]{\endgroup#1\@@endlink}%
\providecommand \@sanitize@url [0]{\catcode `\\12\catcode `\$12\catcode
  `\&12\catcode `\#12\catcode `\^12\catcode `\_12\catcode `\%12\relax}%
\providecommand \@@startlink[1]{}%
\providecommand \@@endlink[0]{}%
\providecommand \url  [0]{\begingroup\@sanitize@url \@url }%
\providecommand \@url [1]{\endgroup\@href {#1}{\urlprefix }}%
\providecommand \urlprefix  [0]{URL }%
\providecommand \Eprint [0]{\href }%
\providecommand \doibase [0]{https://doi.org/}%
\providecommand \selectlanguage [0]{\@gobble}%
\providecommand \bibinfo  [0]{\@secondoftwo}%
\providecommand \bibfield  [0]{\@secondoftwo}%
\providecommand \translation [1]{[#1]}%
\providecommand \BibitemOpen [0]{}%
\providecommand \bibitemStop [0]{}%
\providecommand \bibitemNoStop [0]{.\EOS\space}%
\providecommand \EOS [0]{\spacefactor3000\relax}%
\providecommand \BibitemShut  [1]{\csname bibitem#1\endcsname}%
\let\auto@bib@innerbib\@empty
%</preamble>
\bibitem [{\citenamefont {Cai}\ and\ \citenamefont
  {Plenio}(2013)}]{cai2013chemical}%
  \BibitemOpen
  \bibfield  {author} {\bibinfo {author} {\bibfnamefont {J.}~\bibnamefont
  {Cai}}\ and\ \bibinfo {author} {\bibfnamefont {M.~B.}\ \bibnamefont
  {Plenio}},\ }\bibfield  {title} {\bibinfo {title} {Chemical compass model for
  avian magnetoreception as a quantum coherent device},\ }\href@noop {}
  {\bibfield  {journal} {\bibinfo  {journal} {Physical review letters}\
  }\textbf {\bibinfo {volume} {111}},\ \bibinfo {pages} {230503} (\bibinfo
  {year} {2013})}\BibitemShut {NoStop}%
\bibitem [{\citenamefont {Kominis}(2020)}]{kominis2020quantum}%
  \BibitemOpen
  \bibfield  {author} {\bibinfo {author} {\bibfnamefont {I.}~\bibnamefont
  {Kominis}},\ }\bibfield  {title} {\bibinfo {title} {Quantum relative entropy
  shows singlet-triplet coherence is a resource in the radical-pair mechanism
  of biological magnetic sensing},\ }\href@noop {} {\bibfield  {journal}
  {\bibinfo  {journal} {Physical Review Research}\ }\textbf {\bibinfo {volume}
  {2}},\ \bibinfo {pages} {023206} (\bibinfo {year} {2020})}\BibitemShut
  {NoStop}%
\bibitem [{\citenamefont {Jain}\ \emph {et~al.}(2021)\citenamefont {Jain},
  \citenamefont {Poonia}, \citenamefont {Saha}, \citenamefont {Saha},\ and\
  \citenamefont {Ganguly}}]{jain2021avian}%
  \BibitemOpen
  \bibfield  {author} {\bibinfo {author} {\bibfnamefont {R.}~\bibnamefont
  {Jain}}, \bibinfo {author} {\bibfnamefont {V.~S.}\ \bibnamefont {Poonia}},
  \bibinfo {author} {\bibfnamefont {K.}~\bibnamefont {Saha}}, \bibinfo {author}
  {\bibfnamefont {D.}~\bibnamefont {Saha}},\ and\ \bibinfo {author}
  {\bibfnamefont {S.}~\bibnamefont {Ganguly}},\ }\bibfield  {title} {\bibinfo
  {title} {The avian compass can be sensitive even without sustained electron
  spin coherence},\ }\href@noop {} {\bibfield  {journal} {\bibinfo  {journal}
  {Proceedings of the Royal Society A}\ }\textbf {\bibinfo {volume} {477}},\
  \bibinfo {pages} {20200778} (\bibinfo {year} {2021})}\BibitemShut {NoStop}%
\bibitem [{\citenamefont {Gauger}\ \emph {et~al.}(2011)\citenamefont {Gauger},
  \citenamefont {Rieper}, \citenamefont {Morton}, \citenamefont {Benjamin},\
  and\ \citenamefont {Vedral}}]{gauger2011sustained}%
  \BibitemOpen
  \bibfield  {author} {\bibinfo {author} {\bibfnamefont {E.~M.}\ \bibnamefont
  {Gauger}}, \bibinfo {author} {\bibfnamefont {E.}~\bibnamefont {Rieper}},
  \bibinfo {author} {\bibfnamefont {J.~J.}\ \bibnamefont {Morton}}, \bibinfo
  {author} {\bibfnamefont {S.~C.}\ \bibnamefont {Benjamin}},\ and\ \bibinfo
  {author} {\bibfnamefont {V.}~\bibnamefont {Vedral}},\ }\bibfield  {title}
  {\bibinfo {title} {Sustained quantum coherence and entanglement in the avian
  compass},\ }\href@noop {} {\bibfield  {journal} {\bibinfo  {journal}
  {Physical review letters}\ }\textbf {\bibinfo {volume} {106}},\ \bibinfo
  {pages} {040503} (\bibinfo {year} {2011})}\BibitemShut {NoStop}%
\bibitem [{\citenamefont {Ritz}\ \emph {et~al.}(2000)\citenamefont {Ritz},
  \citenamefont {Adem},\ and\ \citenamefont {Schulten}}]{ritz2000model}%
  \BibitemOpen
  \bibfield  {author} {\bibinfo {author} {\bibfnamefont {T.}~\bibnamefont
  {Ritz}}, \bibinfo {author} {\bibfnamefont {S.}~\bibnamefont {Adem}},\ and\
  \bibinfo {author} {\bibfnamefont {K.}~\bibnamefont {Schulten}},\ }\bibfield
  {title} {\bibinfo {title} {A model for photoreceptor-based magnetoreception
  in birds},\ }\href@noop {} {\bibfield  {journal} {\bibinfo  {journal}
  {Biophysical journal}\ }\textbf {\bibinfo {volume} {78}},\ \bibinfo {pages}
  {707} (\bibinfo {year} {2000})}\BibitemShut {NoStop}%
\bibitem [{\citenamefont {Hore}\ and\ \citenamefont
  {Mouritsen}(2016)}]{hore2016radical}%
  \BibitemOpen
  \bibfield  {author} {\bibinfo {author} {\bibfnamefont {P.~J.}\ \bibnamefont
  {Hore}}\ and\ \bibinfo {author} {\bibfnamefont {H.}~\bibnamefont
  {Mouritsen}},\ }\bibfield  {title} {\bibinfo {title} {The radical-pair
  mechanism of magnetoreception},\ }\href@noop {} {\bibfield  {journal}
  {\bibinfo  {journal} {Annual review of biophysics}\ }\textbf {\bibinfo
  {volume} {45}} (\bibinfo {year} {2016})}\BibitemShut {NoStop}%
\bibitem [{\citenamefont {Smith}\ \emph
  {et~al.}(2022{\natexlab{a}})\citenamefont {Smith}, \citenamefont {Deviers},\
  and\ \citenamefont {Kattnig}}]{smith2022observations}%
  \BibitemOpen
  \bibfield  {author} {\bibinfo {author} {\bibfnamefont {L.~D.}\ \bibnamefont
  {Smith}}, \bibinfo {author} {\bibfnamefont {J.}~\bibnamefont {Deviers}},\
  and\ \bibinfo {author} {\bibfnamefont {D.~R.}\ \bibnamefont {Kattnig}},\
  }\bibfield  {title} {\bibinfo {title} {Observations about utilitarian
  coherence in the avian compass},\ }\href@noop {} {\bibfield  {journal}
  {\bibinfo  {journal} {Scientific reports}\ }\textbf {\bibinfo {volume}
  {12}},\ \bibinfo {pages} {1} (\bibinfo {year}
  {2022}{\natexlab{a}})}\BibitemShut {NoStop}%
\bibitem [{\citenamefont {Smith}\ \emph
  {et~al.}(2022{\natexlab{b}})\citenamefont {Smith}, \citenamefont {Chowdhury},
  \citenamefont {Peasgood}, \citenamefont {Dawkins},\ and\ \citenamefont
  {Kattnig}}]{smith2022driven}%
  \BibitemOpen
  \bibfield  {author} {\bibinfo {author} {\bibfnamefont {L.~D.}\ \bibnamefont
  {Smith}}, \bibinfo {author} {\bibfnamefont {F.~T.}\ \bibnamefont
  {Chowdhury}}, \bibinfo {author} {\bibfnamefont {I.}~\bibnamefont {Peasgood}},
  \bibinfo {author} {\bibfnamefont {N.}~\bibnamefont {Dawkins}},\ and\ \bibinfo
  {author} {\bibfnamefont {D.~R.}\ \bibnamefont {Kattnig}},\ }\bibfield
  {title} {\bibinfo {title} {Driven radical motion enhances cryptochrome
  magnetoreception: Toward live quantum sensing},\ }\href@noop {} {\bibfield
  {journal} {\bibinfo  {journal} {The Journal of Physical Chemistry Letters}\
  }\textbf {\bibinfo {volume} {13}},\ \bibinfo {pages} {10500} (\bibinfo {year}
  {2022}{\natexlab{b}})}\BibitemShut {NoStop}%
\bibitem [{\citenamefont {Hiscock}(2018)}]{hiscock2018long}%
  \BibitemOpen
  \bibfield  {author} {\bibinfo {author} {\bibfnamefont {H.}~\bibnamefont
  {Hiscock}},\ }\emph {\bibinfo {title} {Long-lived spin coherence in radical
  pair compass magnetoreception}},\ \href@noop {} {Ph.D. thesis},\ \bibinfo
  {school} {University of Oxford} (\bibinfo {year} {2018})\BibitemShut
  {NoStop}%
\bibitem [{\citenamefont {Hiscock}\ \emph {et~al.}(2016)\citenamefont
  {Hiscock}, \citenamefont {Worster}, \citenamefont {Kattnig}, \citenamefont
  {Steers}, \citenamefont {Jin}, \citenamefont {Manolopoulos}, \citenamefont
  {Mouritsen},\ and\ \citenamefont {Hore}}]{hiscock2016quantum}%
  \BibitemOpen
  \bibfield  {author} {\bibinfo {author} {\bibfnamefont {H.~G.}\ \bibnamefont
  {Hiscock}}, \bibinfo {author} {\bibfnamefont {S.}~\bibnamefont {Worster}},
  \bibinfo {author} {\bibfnamefont {D.~R.}\ \bibnamefont {Kattnig}}, \bibinfo
  {author} {\bibfnamefont {C.}~\bibnamefont {Steers}}, \bibinfo {author}
  {\bibfnamefont {Y.}~\bibnamefont {Jin}}, \bibinfo {author} {\bibfnamefont
  {D.~E.}\ \bibnamefont {Manolopoulos}}, \bibinfo {author} {\bibfnamefont
  {H.}~\bibnamefont {Mouritsen}},\ and\ \bibinfo {author} {\bibfnamefont
  {P.}~\bibnamefont {Hore}},\ }\bibfield  {title} {\bibinfo {title} {The
  quantum needle of the avian magnetic compass},\ }\href@noop {} {\bibfield
  {journal} {\bibinfo  {journal} {Proceedings of the National Academy of
  Sciences}\ }\textbf {\bibinfo {volume} {113}},\ \bibinfo {pages} {4634}
  (\bibinfo {year} {2016})}\BibitemShut {NoStop}%
\bibitem [{\citenamefont {Rodgers}\ and\ \citenamefont
  {Hore}(2009)}]{rodgers2009chemical}%
  \BibitemOpen
  \bibfield  {author} {\bibinfo {author} {\bibfnamefont {C.~T.}\ \bibnamefont
  {Rodgers}}\ and\ \bibinfo {author} {\bibfnamefont {P.~J.}\ \bibnamefont
  {Hore}},\ }\bibfield  {title} {\bibinfo {title} {Chemical magnetoreception in
  birds: the radical pair mechanism},\ }\href@noop {} {\bibfield  {journal}
  {\bibinfo  {journal} {Proceedings of the National Academy of Sciences}\
  }\textbf {\bibinfo {volume} {106}},\ \bibinfo {pages} {353} (\bibinfo {year}
  {2009})}\BibitemShut {NoStop}%
\bibitem [{\citenamefont {Dalum}\ and\ \citenamefont
  {Hedeg{\aa}rd}(2019)}]{dalum2019theory}%
  \BibitemOpen
  \bibfield  {author} {\bibinfo {author} {\bibfnamefont {S.}~\bibnamefont
  {Dalum}}\ and\ \bibinfo {author} {\bibfnamefont {P.}~\bibnamefont
  {Hedeg{\aa}rd}},\ }\bibfield  {title} {\bibinfo {title} {Theory of chiral
  induced spin selectivity},\ }\href@noop {} {\bibfield  {journal} {\bibinfo
  {journal} {Nano letters}\ }\textbf {\bibinfo {volume} {19}},\ \bibinfo
  {pages} {5253} (\bibinfo {year} {2019})}\BibitemShut {NoStop}%
\bibitem [{\citenamefont {Michaeli}\ and\ \citenamefont
  {Naaman}(2019)}]{michaeli2019origin}%
  \BibitemOpen
  \bibfield  {author} {\bibinfo {author} {\bibfnamefont {K.}~\bibnamefont
  {Michaeli}}\ and\ \bibinfo {author} {\bibfnamefont {R.}~\bibnamefont
  {Naaman}},\ }\bibfield  {title} {\bibinfo {title} {Origin of spin-dependent
  tunneling through chiral molecules},\ }\href@noop {} {\bibfield  {journal}
  {\bibinfo  {journal} {The Journal of Physical Chemistry C}\ }\textbf
  {\bibinfo {volume} {123}},\ \bibinfo {pages} {17043} (\bibinfo {year}
  {2019})}\BibitemShut {NoStop}%
\bibitem [{\citenamefont {Matityahu}\ \emph {et~al.}(2016)\citenamefont
  {Matityahu}, \citenamefont {Utsumi}, \citenamefont {Aharony}, \citenamefont
  {Entin-Wohlman},\ and\ \citenamefont {Balseiro}}]{matityahu2016spin}%
  \BibitemOpen
  \bibfield  {author} {\bibinfo {author} {\bibfnamefont {S.}~\bibnamefont
  {Matityahu}}, \bibinfo {author} {\bibfnamefont {Y.}~\bibnamefont {Utsumi}},
  \bibinfo {author} {\bibfnamefont {A.}~\bibnamefont {Aharony}}, \bibinfo
  {author} {\bibfnamefont {O.}~\bibnamefont {Entin-Wohlman}},\ and\ \bibinfo
  {author} {\bibfnamefont {C.~A.}\ \bibnamefont {Balseiro}},\ }\bibfield
  {title} {\bibinfo {title} {Spin-dependent transport through a chiral molecule
  in the presence of spin-orbit interaction and nonunitary effects},\
  }\href@noop {} {\bibfield  {journal} {\bibinfo  {journal} {Physical Review
  B}\ }\textbf {\bibinfo {volume} {93}},\ \bibinfo {pages} {075407} (\bibinfo
  {year} {2016})}\BibitemShut {NoStop}%
\bibitem [{\citenamefont {G{\"o}hler}\ \emph {et~al.}(2011)\citenamefont
  {G{\"o}hler}, \citenamefont {Hamelbeck}, \citenamefont {Markus},
  \citenamefont {Kettner}, \citenamefont {Hanne}, \citenamefont {Vager},
  \citenamefont {Naaman},\ and\ \citenamefont {Zacharias}}]{gohler2011spin}%
  \BibitemOpen
  \bibfield  {author} {\bibinfo {author} {\bibfnamefont {B.}~\bibnamefont
  {G{\"o}hler}}, \bibinfo {author} {\bibfnamefont {V.}~\bibnamefont
  {Hamelbeck}}, \bibinfo {author} {\bibfnamefont {T.}~\bibnamefont {Markus}},
  \bibinfo {author} {\bibfnamefont {M.}~\bibnamefont {Kettner}}, \bibinfo
  {author} {\bibfnamefont {G.}~\bibnamefont {Hanne}}, \bibinfo {author}
  {\bibfnamefont {Z.}~\bibnamefont {Vager}}, \bibinfo {author} {\bibfnamefont
  {R.}~\bibnamefont {Naaman}},\ and\ \bibinfo {author} {\bibfnamefont
  {H.}~\bibnamefont {Zacharias}},\ }\bibfield  {title} {\bibinfo {title} {Spin
  selectivity in electron transmission through self-assembled monolayers of
  double-stranded dna},\ }\href@noop {} {\bibfield  {journal} {\bibinfo
  {journal} {Science}\ }\textbf {\bibinfo {volume} {331}},\ \bibinfo {pages}
  {894} (\bibinfo {year} {2011})}\BibitemShut {NoStop}%
\bibitem [{\citenamefont {Naaman}\ and\ \citenamefont
  {Waldeck}(2012)}]{naaman2012chiral}%
  \BibitemOpen
  \bibfield  {author} {\bibinfo {author} {\bibfnamefont {R.}~\bibnamefont
  {Naaman}}\ and\ \bibinfo {author} {\bibfnamefont {D.~H.}\ \bibnamefont
  {Waldeck}},\ }\bibfield  {title} {\bibinfo {title} {Chiral-induced spin
  selectivity effect},\ }\href@noop {} {\bibfield  {journal} {\bibinfo
  {journal} {The journal of physical chemistry letters}\ }\textbf {\bibinfo
  {volume} {3}},\ \bibinfo {pages} {2178} (\bibinfo {year} {2012})}\BibitemShut
  {NoStop}%
\bibitem [{\citenamefont {Naaman}\ and\ \citenamefont
  {Waldeck}(2015)}]{naaman2015spintronics}%
  \BibitemOpen
  \bibfield  {author} {\bibinfo {author} {\bibfnamefont {R.}~\bibnamefont
  {Naaman}}\ and\ \bibinfo {author} {\bibfnamefont {D.~H.}\ \bibnamefont
  {Waldeck}},\ }\bibfield  {title} {\bibinfo {title} {Spintronics and
  chirality: Spin selectivity in electron transport through chiral molecules},\
  }\href@noop {} {\bibfield  {journal} {\bibinfo  {journal} {Annu. Rev. Phys.
  Chem}\ }\textbf {\bibinfo {volume} {66}},\ \bibinfo {pages} {263} (\bibinfo
  {year} {2015})}\BibitemShut {NoStop}%
\bibitem [{\citenamefont {Xu}\ \emph {et~al.}(2021)\citenamefont {Xu},
  \citenamefont {Jarocha}, \citenamefont {Zollitsch}, \citenamefont
  {Konowalczyk}, \citenamefont {Henbest}, \citenamefont {Richert},
  \citenamefont {Golesworthy}, \citenamefont {Schmidt}, \citenamefont
  {D{\'e}jean}, \citenamefont {Sowood} \emph {et~al.}}]{xu2021magnetic}%
  \BibitemOpen
  \bibfield  {author} {\bibinfo {author} {\bibfnamefont {J.}~\bibnamefont
  {Xu}}, \bibinfo {author} {\bibfnamefont {L.~E.}\ \bibnamefont {Jarocha}},
  \bibinfo {author} {\bibfnamefont {T.}~\bibnamefont {Zollitsch}}, \bibinfo
  {author} {\bibfnamefont {M.}~\bibnamefont {Konowalczyk}}, \bibinfo {author}
  {\bibfnamefont {K.~B.}\ \bibnamefont {Henbest}}, \bibinfo {author}
  {\bibfnamefont {S.}~\bibnamefont {Richert}}, \bibinfo {author} {\bibfnamefont
  {M.~J.}\ \bibnamefont {Golesworthy}}, \bibinfo {author} {\bibfnamefont
  {J.}~\bibnamefont {Schmidt}}, \bibinfo {author} {\bibfnamefont
  {V.}~\bibnamefont {D{\'e}jean}}, \bibinfo {author} {\bibfnamefont {D.~J.}\
  \bibnamefont {Sowood}}, \emph {et~al.},\ }\bibfield  {title} {\bibinfo
  {title} {Magnetic sensitivity of cryptochrome 4 from a migratory songbird},\
  }\href@noop {} {\bibfield  {journal} {\bibinfo  {journal} {Nature}\ }\textbf
  {\bibinfo {volume} {594}},\ \bibinfo {pages} {535} (\bibinfo {year}
  {2021})}\BibitemShut {NoStop}%
\bibitem [{\citenamefont {Wong}\ \emph {et~al.}(2021)\citenamefont {Wong},
  \citenamefont {Wei}, \citenamefont {Mouritsen}, \citenamefont {Solov'yov},\
  and\ \citenamefont {Hore}}]{wong2021cryptochrome}%
  \BibitemOpen
  \bibfield  {author} {\bibinfo {author} {\bibfnamefont {S.~Y.}\ \bibnamefont
  {Wong}}, \bibinfo {author} {\bibfnamefont {Y.}~\bibnamefont {Wei}}, \bibinfo
  {author} {\bibfnamefont {H.}~\bibnamefont {Mouritsen}}, \bibinfo {author}
  {\bibfnamefont {I.~A.}\ \bibnamefont {Solov'yov}},\ and\ \bibinfo {author}
  {\bibfnamefont {P.}~\bibnamefont {Hore}},\ }\bibfield  {title} {\bibinfo
  {title} {Cryptochrome magnetoreception: four tryptophans could be better than
  three},\ }\href@noop {} {\bibfield  {journal} {\bibinfo  {journal} {Journal
  of the Royal Society Interface}\ }\textbf {\bibinfo {volume} {18}},\ \bibinfo
  {pages} {20210601} (\bibinfo {year} {2021})}\BibitemShut {NoStop}%
\bibitem [{\citenamefont {Fay}(2021)}]{fay2021chirality}%
  \BibitemOpen
  \bibfield  {author} {\bibinfo {author} {\bibfnamefont {T.~P.}\ \bibnamefont
  {Fay}},\ }\bibfield  {title} {\bibinfo {title} {Chirality-induced spin
  coherence in electron transfer reactions},\ }\href@noop {} {\bibfield
  {journal} {\bibinfo  {journal} {The Journal of Physical Chemistry Letters}\
  }\textbf {\bibinfo {volume} {12}},\ \bibinfo {pages} {1407} (\bibinfo {year}
  {2021})}\BibitemShut {NoStop}%
\bibitem [{\citenamefont {Luo}\ and\ \citenamefont
  {Hore}(2021)}]{luo2021chiral}%
  \BibitemOpen
  \bibfield  {author} {\bibinfo {author} {\bibfnamefont {J.}~\bibnamefont
  {Luo}}\ and\ \bibinfo {author} {\bibfnamefont {P.}~\bibnamefont {Hore}},\
  }\bibfield  {title} {\bibinfo {title} {Chiral-induced spin selectivity in the
  formation and recombination of radical pairs: cryptochrome magnetoreception
  and epr detection},\ }\href@noop {} {\bibfield  {journal} {\bibinfo
  {journal} {New Journal of Physics}\ }\textbf {\bibinfo {volume} {23}},\
  \bibinfo {pages} {043032} (\bibinfo {year} {2021})}\BibitemShut {NoStop}%
\bibitem [{\citenamefont {Baumgratz}\ \emph {et~al.}(2014)\citenamefont
  {Baumgratz}, \citenamefont {Cramer},\ and\ \citenamefont
  {Plenio}}]{baumgratz2014quantifying}%
  \BibitemOpen
  \bibfield  {author} {\bibinfo {author} {\bibfnamefont {T.}~\bibnamefont
  {Baumgratz}}, \bibinfo {author} {\bibfnamefont {M.}~\bibnamefont {Cramer}},\
  and\ \bibinfo {author} {\bibfnamefont {M.~B.}\ \bibnamefont {Plenio}},\
  }\bibfield  {title} {\bibinfo {title} {Quantifying coherence},\ }\href@noop
  {} {\bibfield  {journal} {\bibinfo  {journal} {Physical review letters}\
  }\textbf {\bibinfo {volume} {113}},\ \bibinfo {pages} {140401} (\bibinfo
  {year} {2014})}\BibitemShut {NoStop}%
\bibitem [{\citenamefont {Streltsov}\ \emph {et~al.}(2017)\citenamefont
  {Streltsov}, \citenamefont {Adesso},\ and\ \citenamefont
  {Plenio}}]{streltsov2017colloquium}%
  \BibitemOpen
  \bibfield  {author} {\bibinfo {author} {\bibfnamefont {A.}~\bibnamefont
  {Streltsov}}, \bibinfo {author} {\bibfnamefont {G.}~\bibnamefont {Adesso}},\
  and\ \bibinfo {author} {\bibfnamefont {M.~B.}\ \bibnamefont {Plenio}},\
  }\bibfield  {title} {\bibinfo {title} {Colloquium: Quantum coherence as a
  resource},\ }\href@noop {} {\bibfield  {journal} {\bibinfo  {journal}
  {Reviews of Modern Physics}\ }\textbf {\bibinfo {volume} {89}},\ \bibinfo
  {pages} {041003} (\bibinfo {year} {2017})}\BibitemShut {NoStop}%
\bibitem [{\citenamefont {Winter}\ and\ \citenamefont
  {Yang}(2016)}]{winter2016operational}%
  \BibitemOpen
  \bibfield  {author} {\bibinfo {author} {\bibfnamefont {A.}~\bibnamefont
  {Winter}}\ and\ \bibinfo {author} {\bibfnamefont {D.}~\bibnamefont {Yang}},\
  }\bibfield  {title} {\bibinfo {title} {Operational resource theory of
  coherence},\ }\href@noop {} {\bibfield  {journal} {\bibinfo  {journal}
  {Physical review letters}\ }\textbf {\bibinfo {volume} {116}},\ \bibinfo
  {pages} {120404} (\bibinfo {year} {2016})}\BibitemShut {NoStop}%
\bibitem [{\citenamefont {Katsoprinakis}\ \emph {et~al.}(2010)\citenamefont
  {Katsoprinakis}, \citenamefont {Dellis},\ and\ \citenamefont
  {Kominis}}]{katsoprinakis2010coherent}%
  \BibitemOpen
  \bibfield  {author} {\bibinfo {author} {\bibfnamefont {G.}~\bibnamefont
  {Katsoprinakis}}, \bibinfo {author} {\bibfnamefont {A.}~\bibnamefont
  {Dellis}},\ and\ \bibinfo {author} {\bibfnamefont {I.}~\bibnamefont
  {Kominis}},\ }\bibfield  {title} {\bibinfo {title} {Coherent triplet
  excitation suppresses the heading error of the avian compass},\ }\href@noop
  {} {\bibfield  {journal} {\bibinfo  {journal} {New Journal of Physics}\
  }\textbf {\bibinfo {volume} {12}},\ \bibinfo {pages} {085016} (\bibinfo
  {year} {2010})}\BibitemShut {NoStop}%
\bibitem [{\citenamefont {Tiwari}\ and\ \citenamefont
  {Poonia}(2022)}]{https://doi.org/10.48550/arxiv.2209.00736}%
  \BibitemOpen
  \bibfield  {author} {\bibinfo {author} {\bibfnamefont {Y.}~\bibnamefont
  {Tiwari}}\ and\ \bibinfo {author} {\bibfnamefont {V.~S.}\ \bibnamefont
  {Poonia}},\ }\href {https://doi.org/10.48550/ARXIV.2209.00736} {\bibinfo
  {title} {Role of ciss in the radical pair model of avian magnetoreception}}
  (\bibinfo {year} {2022})\BibitemShut {NoStop}%
\bibitem [{\citenamefont {Cintolesi}\ \emph {et~al.}(2003)\citenamefont
  {Cintolesi}, \citenamefont {Ritz}, \citenamefont {Kay}, \citenamefont
  {Timmel},\ and\ \citenamefont {Hore}}]{cintolesi2003anisotropic}%
  \BibitemOpen
  \bibfield  {author} {\bibinfo {author} {\bibfnamefont {F.}~\bibnamefont
  {Cintolesi}}, \bibinfo {author} {\bibfnamefont {T.}~\bibnamefont {Ritz}},
  \bibinfo {author} {\bibfnamefont {C.}~\bibnamefont {Kay}}, \bibinfo {author}
  {\bibfnamefont {C.}~\bibnamefont {Timmel}},\ and\ \bibinfo {author}
  {\bibfnamefont {P.}~\bibnamefont {Hore}},\ }\bibfield  {title} {\bibinfo
  {title} {Anisotropic recombination of an immobilized photoinduced radical
  pair in a 50-$\mu$t magnetic field: a model avian photomagnetoreceptor},\
  }\href@noop {} {\bibfield  {journal} {\bibinfo  {journal} {Chemical Physics}\
  }\textbf {\bibinfo {volume} {294}},\ \bibinfo {pages} {385} (\bibinfo {year}
  {2003})}\BibitemShut {NoStop}%
\bibitem [{\citenamefont {Fay}\ \emph {et~al.}(2020)\citenamefont {Fay},
  \citenamefont {Lindoy}, \citenamefont {Manolopoulos},\ and\ \citenamefont
  {Hore}}]{fay2020quantum}%
  \BibitemOpen
  \bibfield  {author} {\bibinfo {author} {\bibfnamefont {T.~P.}\ \bibnamefont
  {Fay}}, \bibinfo {author} {\bibfnamefont {L.~P.}\ \bibnamefont {Lindoy}},
  \bibinfo {author} {\bibfnamefont {D.~E.}\ \bibnamefont {Manolopoulos}},\ and\
  \bibinfo {author} {\bibfnamefont {P.}~\bibnamefont {Hore}},\ }\bibfield
  {title} {\bibinfo {title} {How quantum is radical pair magnetoreception?},\
  }\href@noop {} {\bibfield  {journal} {\bibinfo  {journal} {Faraday
  discussions}\ }\textbf {\bibinfo {volume} {221}},\ \bibinfo {pages} {77}
  (\bibinfo {year} {2020})}\BibitemShut {NoStop}%
\bibitem [{\citenamefont {Efimova}\ and\ \citenamefont
  {Hore}(2008)}]{efimova2008role}%
  \BibitemOpen
  \bibfield  {author} {\bibinfo {author} {\bibfnamefont {O.}~\bibnamefont
  {Efimova}}\ and\ \bibinfo {author} {\bibfnamefont {P.}~\bibnamefont {Hore}},\
  }\bibfield  {title} {\bibinfo {title} {Role of exchange and dipolar
  interactions in the radical pair model of the avian magnetic compass},\
  }\href@noop {} {\bibfield  {journal} {\bibinfo  {journal} {Biophysical
  Journal}\ }\textbf {\bibinfo {volume} {94}},\ \bibinfo {pages} {1565}
  (\bibinfo {year} {2008})}\BibitemShut {NoStop}%
\end{thebibliography}%

\end{document}